\documentclass[ALICE,manyauthors]{cernphprep}

\newcommand{\pp}           {pp}
\newcommand{\ppbar}        {\mbox{$\mathrm {p\overline{p}}$}}

\newcommand{\pPb}          {\mbox{p--Pb}}

\newcommand{\dAu}          {\mbox{d--Au}}

\newcommand{\dndeta}       {\ensuremath{\mathrm{d}N_\mathrm{ch}/\mathrm{d}\eta}}
\newcommand{\dndetalab}       {\ensuremath{\mathrm{d}N_\mathrm{ch}/\mathrm{d}\eta_\mathrm{lab}}}
\newcommand{\dNdeta}       {\ensuremath{\langle\dndeta\rangle}}

\newcommand{\s}            {\ensuremath{\sqrt{s}}}
\newcommand{\pt}           {\ensuremath{p_{\rm T}}}

\newcommand{\snn}          {\ensuremath{\sqrt{s_{\rm\scriptscriptstyle NN}}}}
\newcommand{\snnbf}      {\ensuremath{\pmb{\sqrt{s_{\rm\scriptscriptstyle NN}}}}}

\newcommand{\Npart}        {\ensuremath{N_\mathrm{part}}}

\newcommand{\avNpart}      {\ensuremath{\langle N_\mathrm{part} \rangle}}
\newcommand{\avNcoll}      {\ensuremath{\langle N_\mathrm{coll} \rangle}}

\newcommand{\Ncoll}        {\ensuremath{N_\mathrm{coll}}}

\newcommand{\dNdetape}     {\ensuremath{\frac{2}{\avNpart}\dNdeta}} 
\newcommand{\dNdetapelab}     {\ensuremath{\frac{2}{\avNpart}{\langle{\mathrm{d}N_\mathrm{ch}/\mathrm{d}\eta_{\mathrm{lab}}}\rangle}}} 
\newcommand{\dNdetaquarklab}     {\ensuremath{\frac{\mu}{{\langle N_\mathrm{q-part} \rangle}}{\langle{\mathrm{d}N_\mathrm{ch}/\mathrm{d}\eta_{\mathrm{lab}}}\rangle}}} 

\usepackage[usenames]{color}
\usepackage{multirow}
\usepackage[normalem]{ulem} 
\usepackage{tabu}
\usepackage{siunitx}
\usepackage{mathtools}
\RequirePackage{color}\definecolor{RED}{rgb}{1,0,0}\definecolor{BLUE}{rgb}{0,0,1}

\usepackage{fancyhdr}
\usepackage[comma,square,numbers,sort&compress,merge]{natbib}
\usepackage{hyperref}
\usepackage{lineno}

\begin{document}%
\newlength{\figlen}
\setlength{\figlen}{0.75\textwidth}

\begin{titlepage}
\PHyear{2018}
\PHnumber{315}      
\PHdate{20 November}  
%

\title{Charged-particle pseudorapidity density at mid-rapidity in \pPb\ collisions at \snnbf\ = 8.16 TeV}
\ShortTitle{\dndeta\ at mid-rapidity in \pPb\ at \snn\ = 8.16 TeV}   

\Collaboration{ALICE Collaboration\thanks{See Appendix~\ref{app:collab} for the list of collaboration members}}
\ShortAuthor{ALICE Collaboration} 

\begin{abstract}
The pseudorapidity density of charged particles, \dndeta , in \pPb\ collisions has been measured at a centre-of-mass energy per nucleon-nucleon pair of \snn\ = 8.16 TeV at mid-pseudorapidity for non-single-diffractive events. 
The results cover 3.6 units of pseudorapidity, $|\eta|<1.8$.
The \dndeta\ value is $19.1\pm0.7$ at $|\eta|<0.5$. 
This quantity divided by \avNpart$/2$ is $4.73\pm0.20$, where \avNpart is the average number of participating nucleons, is  9.5\% higher than the corresponding value for \pPb\ collisions at \snn\ = 5.02 TeV.
Measurements are compared with models based on different mechanisms for particle production.
All models agree within uncertainties with data in the Pb-going side, while HIJING overestimates, showing a symmetric behaviour, and EPOS underestimates the p-going side of the \dndeta\ distribution.
Saturation-based models reproduce the distributions well for $\eta>-1.3$.
The \dndeta\ is also measured for different centrality estimators, based both on the charged-particle multiplicity and on the energy deposited in the Zero-Degree Calorimeters. 
A study of the implications of the large multiplicity fluctuations due to the small number of participants for systems like \pPb\ in the centrality calculation for multiplicity-based estimators is discussed, demonstrating the advantages of determining the centrality with energy deposited near beam rapidity. 
\end{abstract}
\end{titlepage}
\setcounter{page}{2}

%
%

\section{Introduction}
\label{intro}
Particle production in proton--nucleus (pA) collisions is influenced by nuclear effects in the initial state. 
In particular, \pPb\ collisions are a valuable tool to study initial-state effects, which are present as a consequence of the nucleons being bound into nuclei.
Additionally, the particle multiplicity is an important tool to study the various theoretical models of gluon saturation, which contain different treatments of the upper limit in the growth of the parton density. Therefore, pseudorapidity density measurements can provide constraints to the modelling of the initial state at small Bjorken-$x$.
Moreover, evidence for collective phenomena have been observed in \pPb\ collisions, with the magnitude of the effects increasing with event multiplicity~\cite{Abelev:2012ola, Aad:2012gla, CMS:2012qk, Abelev:2013bla, ABELEV:2013wsa, Abelev:2013haa, Abelev:2014mda, Adam:2015bka, Khachatryan:2015waa}.
Proton--nucleus collisions serve as a tool to study also final-state effects that are sensitive to the formation of a Quark--Gluon Plasma in heavy-ion collisions, under active scrutiny by the community~\cite{Nagle:2018nvi}.
For these reasons, it is important to understand the collision geometry and the global properties of the system produced in \pPb\ collisions.

This paper presents a measurement of the primary charged-particle density in \pPb\ collisions, \dndetalab , at a nucleon-nucleon centre-of-mass energy of \snn\ = 8.16 TeV for pseudorapidities $|\eta_\mathrm{lab}|<1.8$ in the laboratory system.
A primary charged particle is defined as a charged particle with a mean proper lifetime $\tau$ larger than 1 cm$/c$, which is either produced directly in the interaction, or from decays of particles with $\tau$ smaller than 1 cm$/c$, excluding particles produced in interactions with the beam pipe, material of the subdetectors, cables and support structures~\cite{ALICE-PUBLIC-2017-005}.
The dominant processes in \pPb\ collisions are the non-diffractive ones. 
Diffractive events can be single-, double- or central-diffractive and results are presented for non-single-diffractive (NSD) events.
Data are compared to other experimental measurements available in pp, \pPb , d--Au and AA collisions.
Results are compared also with simulations (performed with HIJING 2.1~\cite{Deng:2010mv, Xu:2012au}, EPOS 3~\cite{Drescher:2000ha, Werner:2010aa, Werner:2013tya} and EPOS LHC~\cite{Pierog:2013ria}) and calculations incorporating the saturation of the gluon density in the colliding hadrons (MC-rcBK~\cite{Albacete:2010ad, Albacete:2012xq} and KLN~\cite{Kharzeev:2002ei, Dumitru:2011wq}).

The rest of this article is organised in the following way: Sec.~\ref{setup} describes the experimental conditions and the detectors used to measure the centrality of the event and the pseudorapidity density of charged particles. In Sec.~\ref{centrality}, the centrality determination methodologies are described, both the ones using the multiplicity distributions of charged particles and the alternative one that relies on the energy collected in the neutron Zero-Degree Calorimeters (ZDCs). 
Section~\ref{analysis} explains, in detail, the analysis procedure to measure the \dndeta . 
The systematic uncertainties are described in Sec.~\ref{uncertainties}, and the results along with comparisons to models are presented in Sec.~\ref{res}. 
A brief summary and conclusions are given in Sec.~\ref{summary}.

\section{Experimental setup}
\label{setup}

The \pPb\ data were provided by the Large Hadron Collider (LHC) in December 2016.
There were two configurations that were exploited: in one, denoted by \pPb\ below, the proton beam circulated towards the negative $z$ direction in the ALICE laboratory system, while $\prescript{208}{}{\rm Pb}$ ions circulated in the opposite direction; in the second configuration, denoted by Pb--p, the direction of both beams was reversed. The total luminosity was 0.06 nb$^{-1}$, corresponding to around 120 million minimum-bias (MB) events in the \pPb\ and Pb--p configurations.
The beams in both rings have the same magnetic rigidity.
The nucleon-nucleon centre-of-mass energy was \snn\ = 8.16 TeV, with both p and Pb beams at 6.5 TeV per proton charge.
Due to the asymmetric collision system, there is a shift in the centre-of-mass rapidity of \mbox{$\Delta y$ = 0.465} in the direction of the proton beam.

Full details of the ALICE detector are given elsewhere~\cite{Aamodt:2008zz, Abelev:2014ffa}.
The main element used for the analysis was the Silicon Pixel Detector (SPD): the two innermost cylindrical layers of the ALICE Inner Tracking System~\cite{Aamodt:2008zz}, made of hybrid silicon pixel chips.
The SPD is located inside a solenoidal magnet that provides a magnetic field of 0.5 T.
The first layer covers $\vert\eta_{\rm lab}\vert<2.0$ for collisions at the nominal Interaction Point (IP), while the second covers  $\vert\eta_{\rm lab}\vert<1.4$.
The layers have full azimuthal coverage and radii of $3.9$~cm and $7.6$~cm, respectively. In total, the SPD has $9.8\times10^6$ silicon pixels, each of size $50\times425$~${\rm {\mu}m}^2$.

The MB trigger signal is given by a hit in both the V0 hodoscopes~\cite{Abbas:2013taa}. 
The V0 detector  is composed of two arrays of 32 scintillators positioned at 3.3 m (V0A) and -0.90 m (V0C) from the nominal IP along the beam axis.
Each array has a ring structure segmented into 4 radial and 8 azimuthal sectors.
The detector has full azimuthal coverage in the pseudorapidity ranges $2.8<\eta_{\rm lab}<5.1$ and $-3.7<\eta_{\rm lab}<-1.7$.
The signal amplitudes and particle arrival times are recorded for each of the 64 scintillators. 
The V0 is well suited for triggering thanks to its good timing resolution (below 1~ns) and its large angular acceptance.
The timing is used to discriminate the beam--beam collisions from background events, like beam--gas and beam--halo events, produced outside the interaction region.
The neutron ZDCs~\cite{Gallio:381433} are likewise utilised for background rejection.
The neutron calorimeters, ZNs, are quartz-fibre spaghetti calorimeters placed at zero degrees with respect to the LHC beam axis, positioned at 112.5 m (ZNA) and -112.5 m (ZNC) from the nominal IP. 
ZNs detect neutral particles emitted at pseudorapidities $\vert\eta_{\rm lab}\vert>8.7$ and have an energy resolution of around 18\% for neutron energies of 2.56 TeV.
ALICE is equipped also with the proton calorimeters, ZPs, which are not used in the analysis.

A subsample of 6.8 million events is analysed for \pPb\ collisions, with an average number of interactions per bunch crossing, $\langle\mu\rangle$ of 0.004.
A subsample of 2.7 million events is analysed for Pb--p collisions, with $\langle\mu\rangle$ = 0.007.
The comparison of \pPb\ and Pb--p results is used to assess the systematic uncertainties.
The hardware MB trigger is configured to have high efficiency for hadronic events, requiring a signal in both V0A and V0C.
Beam--gas and beam--halo interactions are suppressed in the analysis by requiring offline the arrival time of particles in the V0 and ZN detectors to be compatible with collisions from the nominal IP. 
The contamination from background is estimated to be negligible through control triggers on non-colliding bunches. 

The event sample after trigger and timing selection consisted of NSD, single-diffractive (SD), and electromagnetic (EM) interactions.
The MB trigger efficiency for NSD events is estimated to be 99.2\% using the DPMJet Monte Carlo event generator~\cite{Roesler:2000he}, and 99.5\% using HIJING 1.36~\cite{Wang:1991hta}. 
HIJING 1.36 combines perturbative-QCD processes with soft interactions, and includes a strong impact parameter dependence of parton shadowing.
DPMJet is based on the Gribov-Glauber approach and treats soft and hard scattering processes in a unified way.
It includes incoherent SD collisions of the projectile proton with target nucleons; these interactions are concentrated mainly on the surface of the nucleus.
The generated particles are transported through the experimental setup using the GEANT3~\cite{Brun:1994aa} software package.
SD collisions are removed in DPMJet by requiring that at least one of the binary nucleon-nucleon interactions is NSD.
The SD and EM contaminations are estimated from Monte Carlo simulation studies to be around 0.03\% and below 0.3\%, respectively.  

Among the selected events in data, 99\% had a primary interaction vertex.
In DPMJet this fraction was 99.6\% (99.8\% for HIJING 1.36), with a trigger and selection efficiency for events without a primary vertex of 28\% (23.1\%). 
Taking into account the difference of the fraction of events without a vertex in the data and the simulation, the overall selection efficiency for NSD events in the analysis is estimated to be 97.0\% (96.2\%) according to DPMJet (HIJING 1.36).

\section{Centrality determination}
\label{centrality}
The Glauber model~\cite{Miller:2007ri, Loizides:2017ack} is used to calculate the number of participating nucleons (participants), \Npart , and the corresponding number of nucleon-nucleon collisions, \Ncoll , which depend on the collision impact parameter, $b$.
Indeed, the number of produced particles changes with the variation of the amount of matter overlapping in the collision region; \Npart\ and \Ncoll\ describe quantitatively this variation.
In pA collisions, \mbox{\Ncoll\ $=$ \Npart\ $-1$}.
Using the Glauber model, it is possible to calculate the probability distributions of the relevant parameters, \Npart\ and \Ncoll , which for pA collisions are loosely correlated to $b$.
Centrality classes are defined as percentile intervals of the visible cross section, which determines the event sample after the selections described in Sec.~\ref{setup}.
The number of participating nucleons and nucleon-nucleon collisions are calculated, accordingly, for the visible cross section.

The centrality is determined for three different estimators, two of which are based on observables well separated in pseudorapidity to limit the effect of short-range correlations in the collision region.
The method founded on multiplicity-based estimators is derived by fitting the measured charged-particle multiplicity distributions with an \Ncoll\ distribution obtained from the Glauber model convoluted with a Negative Binomial Distribution (NBD) to model the multiplicity produced in a single collision.
Multiplicity fluctuations play an important role in pA collisions.
The range of multiplicities used to define a centrality class in the case of pA collisions is of the same order of magnitude as the multiplicity fluctuations width~\cite{Adam:2014qja}. 
Therefore, a biased sample of nucleon-nucleon collisions is selected using multiplicity. 
Samples of high-multiplicity events select not only a class with larger than average \avNpart , but also one which is widely spread in \Ncoll\ and that leads to deviations from the scaling of hard processes with Multiple Parton Interactions (MPI). 
These high-multiplicity nucleon-nucleon collisions have a higher particle mean transverse momentum $p_{\rm T}$, and are collisions where MPI are more likely~\cite{Abelev:2013bla}.
The opposite happens for low-multiplicity events.

The centrality determined from the hybrid method, described in Sec.~\ref{centralityZN} using the energy deposited in the ZDCs, on the contrary, minimises biases on the binary scaling of hard processes.
Indeed, the ZDCs detect, at large $\eta$ separation from the central region, the nucleons produced in the interaction through the nuclear de-excitation process or knocked out by participants (called slow nucleons). 
A heuristic approach based on extrapolation from low-energy data is discussed in a previous publication~\cite{Adam:2014qja}.

\subsection{Centrality from charged-particle distributions}

\begin{figure}[t]
\begin{center}
\includegraphics[width=\figlen]{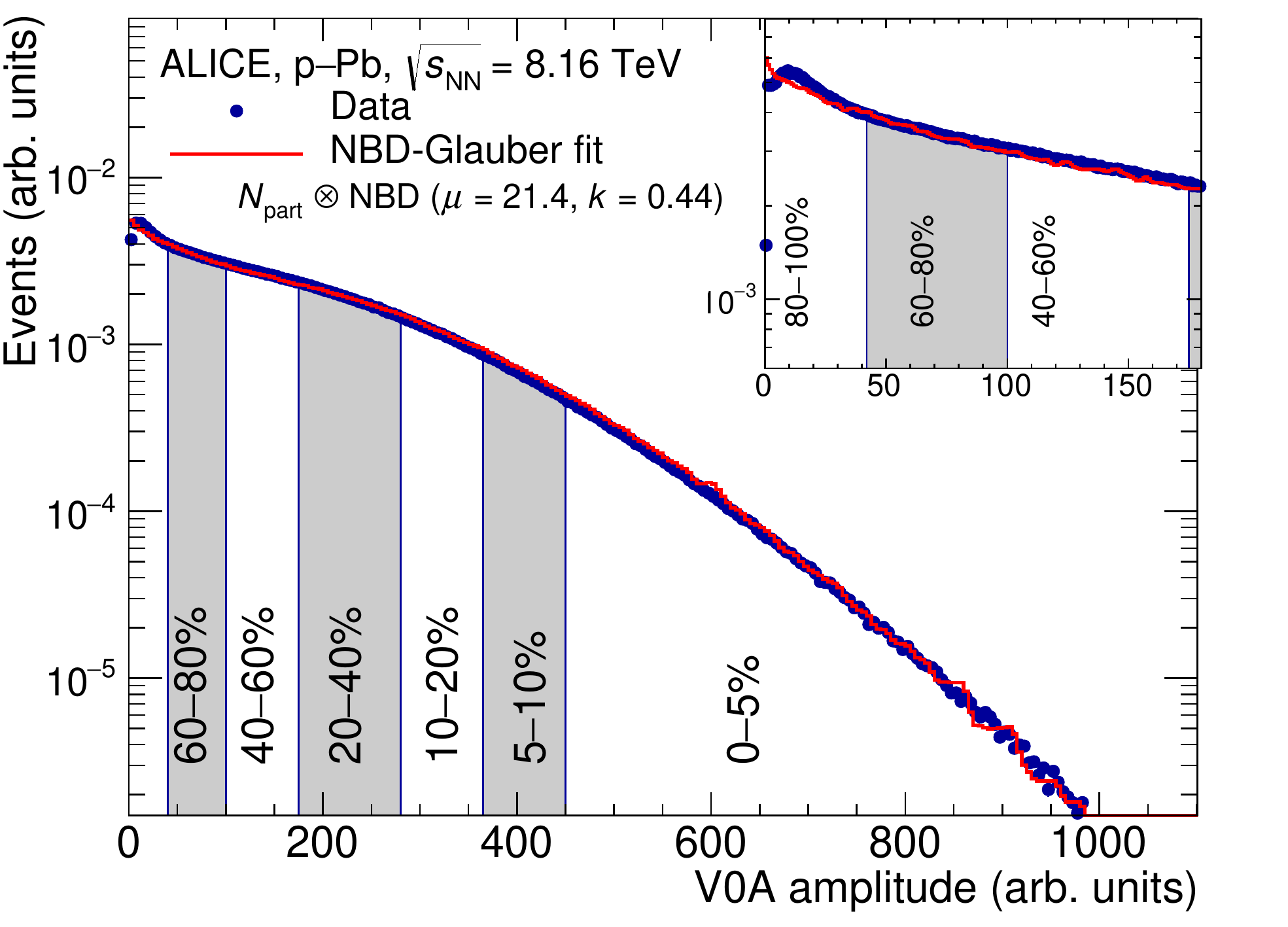}
\caption{Distribution of the sum of amplitudes in V0A (Pb-going side) and the NBD-Glauber fit in red. Centrality classes are indicated by vertical lines and the inset shows the most peripheral events in more detail.}
\label{fig:1}
\end{center}
\end{figure}
\begin{table}[t] \centering
\caption{Mean values of \Npart , \Ncoll\ and $T_{\rm pPb}$ of \pPb\ collisions for MB and centrality classes defined by slices in CL1 and V0A. The values are obtained with a Glauber Monte Carlo calculation coupled to an NBD to fit CL1 and V0A distributions.} \label{tab:NBD-Glauber}
\small
\begin{tabular}{
  @{}
  l
  S[table-format=2.2]
  S[table-format=1.1]
  S[table-format=1.2]
  S[table-format=2.2]
  S[table-format=1.1]
  S[table-format=1.2]
  S[table-format=1.4]  
   S[table-format=1.3]
  S[table-format=1.4]
  @{}
}
\hline
{Centrality (\%)}		& {\avNpart	}		& {RMS}	& {syst.} & {\avNcoll}	& {RMS}	& {syst.}  & {$\langle T_{\rm pPb}\rangle$ (mb$^{-1}$)} & {RMS (mb$^{-1}$)} & {syst. (mb$^{-1}$)}\\ \hline\hline
\SIrange{0}{100}{}		& 8.09		& 5.3	 & 0.17 & 7.09 & 5.3 & 0.16	& 0.0978   & 0.073	& 0.0021 \\ \hline		
\multicolumn{10}{c}{CL1 Estimator}  \\ \hline
\SIrange{0}{5}{}	 		& 17.0 	& 3.6 & 0.6   & 16.0 	& 3.6 & 0.6 	& 0.220 	& 0.050 & 0.008 \\ \hline
\SIrange{5}{10}{}	 	& 15.0 	& 3.5 & 0.4   & 14.0 	& 3.5 & 0.4 	& 0.193   & 0.048 & 0.006 \\ \hline
\SIrange{10}{20}{}	 	& 13.4 	& 3.5 & 0.4   & 12.4 	& 3.5 & 0.4 	& 0.172	 & 0.048 & 0.004 \\ \hline
\SIrange{20}{40}{}	& 10.9 	& 3.6 & 0.2   & 9.9 	& 3.6 & 0.2 	& 0.136 	 & 0.050 & 0.003 \\ \hline
\SIrange{40}{60}{} 	& 7.47 	& 3.3 & 0.15 & 6.47 	& 3.3 & 0.15 	& 0.0893 & 0.046 & 0.0022 \\ \hline
\SIrange{60}{80}{} 	& 4.53 	& 2.4 & 0.09 & 3.53 	& 2.4 & 0.09 	& 0.0487 & 0.033 & 0.0013 \\ \hline
\SIrange{80}{100}{} 	& 2.76 	& 1.2 & 0.03 & 1.76 	& 1.2 & 0.03 	& 0.0242 & 0.016 & 0.0004 \\ \hline
\multicolumn{10}{c}{V0A Estimator}  \\ \hline
\SIrange{0}{5}{} 		& 16.5 	& 3.8 & 0.6   & 15.5 	& 3.8 & 0.6 	& 0.213 	& 0.052 & 0.008 \\ \hline  	
\SIrange{5}{10}{} 	& 14.6 	& 3.7 & 0.4   & 13.6 	& 3.7 & 0.4 	& 0.188   & 0.052 & 0.006 \\ \hline	
\SIrange{10}{20}{} 	& 13.1 	& 3.9 & 0.4   & 12.1 	& 3.9 & 0.4 	& 0.167	 & 0.053 & 0.004 \\ \hline	
\SIrange{20}{40}{} 	& 10.7 	& 4.0 & 0.2   & 9.7 	& 4.0 & 0.2 	& 0.134 	 & 0.055 & 0.003 \\ \hline 
\SIrange{40}{60}{} 	& 7.64 	& 3.7 & 0.16 & 6.64 	& 3.7 & 0.16 	& 0.0916 & 0.051 & 0.0023 \\ \hline 
\SIrange{60}{80}{} 	& 4.80 	& 2.7 & 0.10 & 3.80 	& 2.7 & 0.10 	& 0.0525 & 0.037 & 0.0013 \\ \hline 
\SIrange{80}{100}{} 	& 2.88 	& 1.4 & 0.03 & 1.88 	& 1.4 & 0.03 	& 0.0260 & 0.019 & 0.0004 \\ \hline 
\end{tabular} 
\end{table}

In the method based on multiplicity estimators~\cite{Adam:2014qja}, the events are classified into centrality classes using either the number of clusters in the outer layer of the SPD (CL1 estimator) with acceptance $\eta_{\rm lab}< 1.4$, or the amplitude measured by the V0 in the Pb-remnant side, A-side, for \pPb\ (V0A estimator) or in the C-side for Pb--p (V0C estimator) collisions.
The amplitudes are fitted with a Monte Carlo implementation of the Glauber model assuming that the number of sources is given by the \Npart$/2$ convoluted with an NBD, which is the assumed particle production per source, parametrised with $\mu$ and $k$, where $\mu$ is the mean multiplicity per source and $k$ controls the contribution at high multiplicity. 
The nuclear density for Pb is modelled by a Woods-Saxon distribution for a spherical nucleus with a radius of 6.62 $\pm$ 0.06 fm and a skin thickness of \mbox{0.55 $\pm$ 0.01 fm~\cite{DeJager:1987qc}}.
The hard-sphere exclusion distance between nucleons is 0.40 $\pm$ 0.40 fm. 
For \mbox{\snn\ = 8.16 TeV} collisions, an inelastic nucleon-nucleon cross section of 72.5 $\pm$ 0.5 mb is used, obtained by interpolation of cross section experimental values~\cite{DeJager:1987qc}. 

The measured V0A distribution with the NBD-Glauber fit is shown in Fig.~\ref{fig:1}.
A similar fit has been performed for the CL1 estimator.
The failure of the chosen fit function for amplitudes smaller than about 10 is due to trigger inefficiencies in peripheral collisions.
The average number of participants, collisions and nuclear overlap function, $\langle T_{\rm pPb}\rangle$, are calculated from the NBD-Glauber simulation for every defined centrality class.
The values for the different estimators are given in Tab.~\ref{tab:NBD-Glauber}.
The systematic uncertainties are obtained by repeating the fit, varying the Glauber parameters (radius, skin thickness and hard-sphere exclusion) within their uncertainties.
The number of participants for all selected events is on average  \Npart\ = 8.09 $\pm$ 0.17. 
The increase in the average \Npart , when calculated for NSD collisions only, is of around 2\% and within systematic uncertainties.
The geometrical properties determined with the NBD-Glauber model are robust and approximately independent of the centrality estimator used, within the model assumptions of this approach.

\subsection{Centrality from Zero-Degree Calorimeter and the hybrid method}
\label{centralityZN}

\begin{figure}[t]
\begin{center}
\includegraphics[width=\figlen]{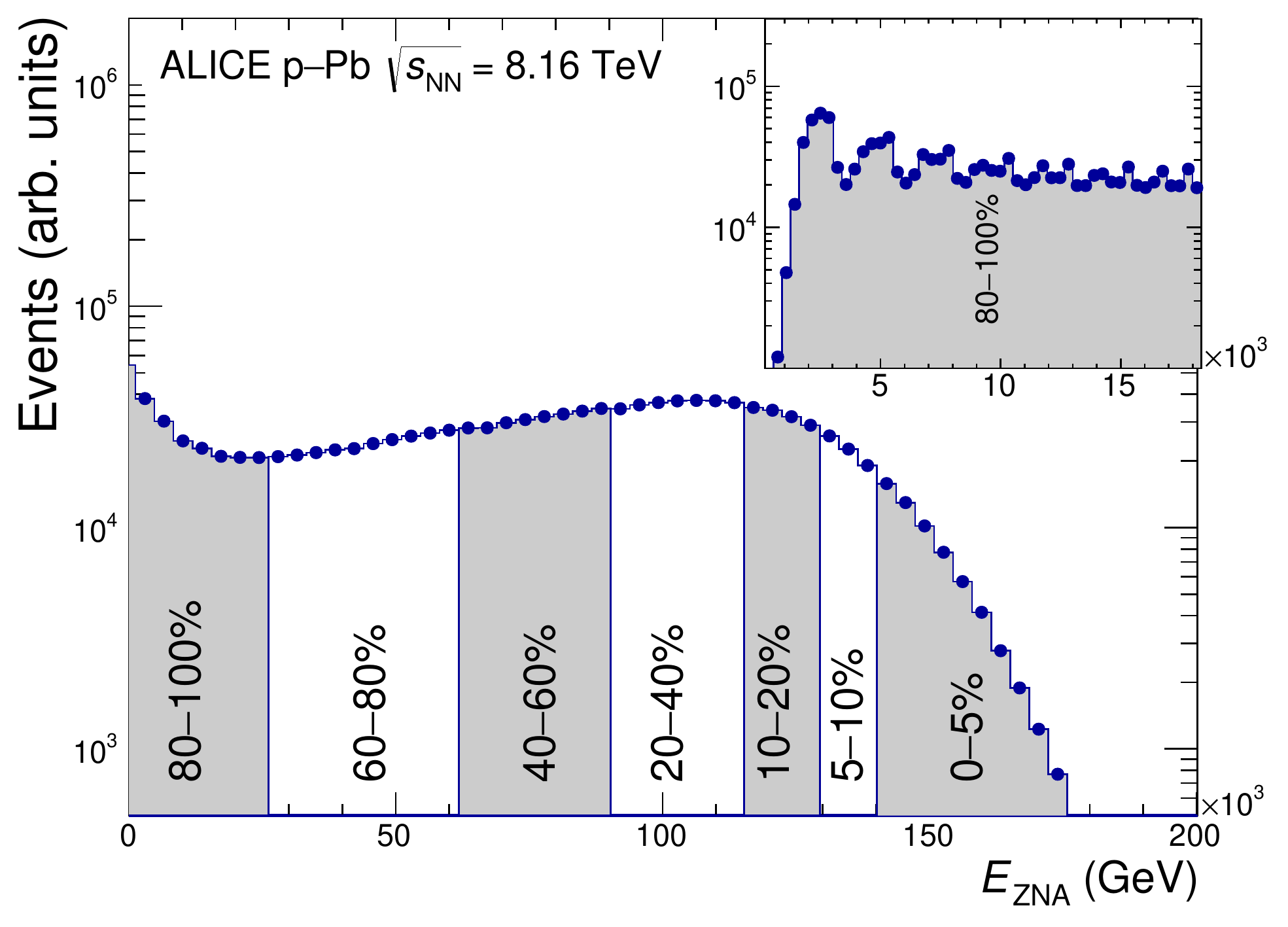}
\caption{Distribution of the neutron energy spectrum measured in the Pb-going side (ZNA). Centrality classes are indicated by vertical lines and the inset shows the most peripheral events in more detail.}
\label{fig:2}
\end{center}
\end{figure}
\begin{table}[t] \centering
\caption{Average number of hadronic nucleon collisions for the ZNA estimator, with the assumption of charged-particle multiplicity at mid-rapidity proportional to \Npart , \avNcoll$^{\rm mult}$, and assuming the signal in V0 proportional to \Ncoll , \avNcoll$^{\text{Pb-side}}$.} \label{tab:Ncoll}
\begin{tabular}{
  @{}
  l
  S[table-format=2.2]
  S[table-format=2.2]
  S[table-format=1.1]
  @{}
} 
\hline
Centrality (\%)		& \avNcoll$^{\rm mult}$  &   \avNcoll$^{\text{Pb-side}}$ &  {syst. (\%)}  \\ \hline\hline
0--5		& 13.4		& 14.2		& 6.4 \\ \hline
5--10		& 12.5		& 12.9		& 3.9 \\ \hline
10--20	& 11.5		& 11.8		& 3.4 \\ \hline
20--40	& 9.81		& 9.77		& 2.3 \\ \hline
40--60	& 7.09		& 6.83		& 4.3 \\ \hline
60--80	& 4.28		& 4.09		& 4.9\\ \hline
80--100	& 2.08		& 2.13		& 3.3 \\ \hline
\end{tabular}
\end{table}

The ZNs detect the slow neutrons produced in the interaction. 
The multiplicity of slow nucleons is monotonically related to \Ncoll , and can, therefore, be used to determine the centrality of the collision~\cite{Adam:2014qja}.
The ZPs are not used, since the uncertainty on \Ncoll\ would be much larger.
The experimental distribution of the neutron energy spectrum measured in the Pb-going side, $E_{\rm ZNA}$, is shown in Fig.~\ref{fig:2} and it is used for the hybrid method, which aims to provide an unbiased centrality estimator.
It is based on two assumptions, the first is that the event selection based on the energy deposited in the ZDCs is free from the multiplicity fluctuation biases in the particle production at mid-rapidity. 
The second assumption is that the wounded nucleon model holds~\cite{Bialas:2007eg} and that some observables, defined below, scale linearly with \Ncoll\ and \Npart\ allowing one to establish a relationship to the collision geometry. 
Two sets of \avNcoll\ are calculated: $N_{\rm coll}^{\rm mult}$ and $N_{\rm  coll}^{\text{Pb-side}}$ for each centrality bin $i$ estimated using ZN.
The first set is computed assuming that the charged-particle multiplicity at mid-rapidity is proportional to the \Npart : \avNpart$^{\rm mult}_{i}=$ \avNpart$_{\rm MB}\cdot(\langle\text{d}N_{\rm ch}/\rm{d}\eta_{lab}\rangle_{i}/\langle \text{d}N_{\rm ch}/\rm{d}\eta_{\rm lab}\rangle_{\rm{MB}})$, where \avNpart$_{\rm MB}$ is the average number of participating nucleons in MB collisions reported in Tab.~\ref{tab:NBD-Glauber}, and, consequently: \avNcoll$_{i}^{\rm mult}=$ \avNpart$^{\rm mult}_{i}-1$.
The second set is calculated using the Pb-side multiplicity: \avNcoll$_{i}^{\text{Pb-side}}=$ \avNcoll$_{\rm MB}\cdot(\langle S\rangle_{i}/\langle S\rangle_{\rm MB})$, where $S$ is the raw signal of the innermost ring of V0A for \pPb\ ($4.5<\eta_{\rm lab}<5.1$) and V0C for Pb--p collisions ($-3.7<\eta_{\rm lab}<-3.2$).
A comparison of the \Ncoll\ values obtained for the various estimators is reported in Tab.~\ref{tab:Ncoll} for \pPb\ collisions.
The two different sets are consistent among each other and with the values calculated for Pb--p.
The systematic uncertainties come from the uncertainty on the \Ncoll\ for 0--100\% in Tab.~\ref{tab:NBD-Glauber} summed with the maximum difference between the $N_{\rm coll}^{\rm mult}$ and $N_{\rm  coll}^{\text{Pb-side}}$.

\section{Analysis procedure}
\label{analysis}

The technique for the \dndetalab\ measurement is the same as the one employed at \snn\ = 5.02 TeV~\cite{ALICE:2012xs, Adam:2014qja}.
The pseudorapidity acceptance in the laboratory system depends on the position of the primary interaction vertex along the beamline, $z_{\rm vtx}$. 
The position of the primary vertex is obtained by correlating hits in the two silicon-pixel layers (SPD vertex).
The selection of a reconstructed vertex within $|z_{\rm vtx}|<15$ cm allows a range of $|\eta_{\rm lab}|<1.8$ to be covered.
In order to maximise the pseudorapidity coverage, instead of tracks we use tracklets (short track segments) formed using two hits in the SPD, one in the first and one in the second layer.
In order to select combinations corresponding to charged particles, the angular difference in the azimuthal direction, $\Delta\varphi$, and in the polar direction, $\Delta\theta$, of the inner and outer layer hit with respect to the reconstructed primary vertex  is determined for each pair of hits.
Afterwards, the sum of the squares of the weighted differences in azimuth and polar angles $\delta^{2}=(\Delta\varphi/\sigma_{\varphi})^{2}+(\Delta\theta/\sigma_{\theta})^{2}$ is required to be less than 1.5, where \mbox{$\sigma_{\varphi}=60$ mrad} and \mbox{$\sigma_{\theta}=25\sin^{2}\theta$ mrad}, where the $\sin^{2}$ factor takes the dependence of the pointing resolution on $\theta$ into account.
With such a requirement, tracklets corresponding to charged particles with \pt\ $>50$ MeV$/c$ are effectively selected.
Particles with lower \pt\ are mostly absorbed by the detector material or lost due to the bending in the magnetic field.
A cross check utilising pp collisions~\cite{Adam:2015gka} has shown full compatibility of analyses using tracklets and tracks, where the tracks have been reconstructed in the Time Projection Chamber matched with clusters in the Inner Tracking System.

The raw multiplicity measured by tracklets needs to be corrected for (i) the acceptance and efficiency of a primary track to be reconstructed as a tracklet, (ii) the contribution from combinatorial tracklets, i.e.~those whose two hits do not originate from the same primary particle, (iii) the difference between the fraction of events without a vertex in the data and in the simulation and (iv) the secondary-particle contamination.
The first three corrections are computed using simulated data from the HIJING 1.36 or DPMJet event generators.
The centrality definition in the simulated data is adjusted such that the particle density is similar to that in real data for the same centrality classes.
The correction factors (i) and (ii), determined as a function of $z$ and $\eta_{\rm lab}$, are on average around 1.5 for the acceptance and reconstruction efficiency, and around 0.02 for the combinatorial background removal in MB and centrality-dependent measurements at mid-rapidity, independently of the estimator selected and the centrality class.
At $|\eta_{\rm lab}|=1.8$ the combinatorial background contribution reaches a maximum value of 0.07.
We further correct the measurement by the difference in the fraction of events without a vertex observed in data and simulation.
The correction for MB \dndetalab\ amounts to 2.2\% (3.4\%) when using DPMJet (HIJING 1.36).
Since the centrality classes are defined as percentiles of the visible cross section, the centrality-dependent measurements are not corrected for the trigger inefficiencies.
Differences in strange-particle content observed at lower beam energies~\cite{Abelev:2013haa, ALICE:2017jyt} have been used for a data-driven correction applied to the generator output, giving rise to a correction factor of $-0.6$\%, independent of centrality.

\section{Systematic uncertainties}
\label{uncertainties}

\begin{table}[t] \centering
\small
    \caption{Overview of the sources of systematic uncertainties.} \label{tab:systematics}
  \begin{tabular}{
    @{}
  l|
  S[table-format=1.1]
  S[table-format=1.1]|
   S[table-format=1.1]
  S[table-format=1.1]|
   S[table-format=1.1]
  S[table-format=1.1]
  @{}
}
\hline             
		& \multicolumn{6}{c}{Uncertainty (\%)} \\ \hline
   Source  &  \multicolumn{2}{c|}{0--100\%} & \multicolumn{2}{c|}{0--5\%} & \multicolumn{2}{c}{80--100\%} \\
    					  & {$\eta=0$} & {$|\eta|=1.8$} & {$\eta=0$} & {$|\eta|=1.8$} & {$\eta=0$} & {$|\eta|=1.8$}   \\ \hline\hline
    Tracklet selection criteria & {negligible} & 0.5 & {negligible} & 0.5  & {negligible} & 0.5 \\
    Weak-decay contamination & \multicolumn{2}{c|}{1.3} & \multicolumn{2}{c|}{1.3} & \multicolumn{2}{c}{1.3} \\
    Detector acceptance and efficiency & \multicolumn{2}{c|}{{2.2}} & \multicolumn{2}{c|}{{2.2}} & {2.2} & {2.8} \\
    Trigger efficiency & \multicolumn{2}{c|}{0.8} & \multicolumn{2}{c|}{--} & \multicolumn{2}{c}{1.7} \\   
    Event-generator dependence & \multicolumn{2}{c|}{1.2} & \multicolumn{2}{c|}{--} & \multicolumn{2}{c}{--}\\ 
    Background subtraction & \multicolumn{2}{c|}{0.3} &  \multicolumn{2}{c|}{0.3} &   \multicolumn{2}{c}{0.3} \\
    Material budget & \multicolumn{2}{c|}{0.1} & \multicolumn{2}{c|}{0.1} & \multicolumn{2}{c}{0.1}\\
    Particle composition & \multicolumn{2}{c|}{0.3} & \multicolumn{2}{c|}{0.3} & \multicolumn{2}{c}{0.3} \\
    Zero-\pt\ extrapolation & \multicolumn{2}{c|}{negligible} &  \multicolumn{2}{c|}{negligible}  &  \multicolumn{2}{c}{negligible} \\
    Pileup & \multicolumn{2}{c|}{negligible} &  \multicolumn{2}{c|}{negligible}  &  \multicolumn{2}{c}{negligible} \\
    \hline\hline
    Total                       & \multicolumn{2}{c|}{3.0} & \multicolumn{2}{c|}{2.6}  & {3.1} &{3.6}  \\ \hline
  \end{tabular}
\end{table}

Several sources of systematic uncertainties were investigated.
The uncertainty coming from the selection of the tracklet quality value  $\delta^{2}$ is negligible at mid-rapidity and amounts to 0.5\% at $|\eta_{\rm lab}|=1.8$. 
The other uncertainties associated to the MB \dndetalab\ are independent of the pseudorapidity.
The uncertainty resulting from the subtraction of the contamination from weak decays of strange hadrons is estimated to be about 1.3\%.
It is estimated by varying the amount of strange particles except kaons by $\pm50$\%.
The uncertainty in detector acceptance and reconstruction efficiency is estimated to be 2.2\% by carrying out the analysis for different slices of the $z_{\rm vtx}$ position distribution and with subsamples in azimuth.
The measurement for Pb--p collisions gives rise to an additional contribution of 1.8\%, when reflected in $\eta_{\rm lab}$, for the most peripheral centrality bins (80--100\%), and 1.1\% for 60--80\% at $|\eta_{\rm lab}|=1.8$, and is added to the systematic uncertainty for acceptance. 
For the other centrality bins and the MB result the difference among p--Pb and Pb--p is negligible and already accounted for in the acceptance and reconstruction efficiency uncertainty.
The uncertainty related to the trigger and event selection efficiency for NSD collisions is estimated to be 0.8\% by taking into account the differences in the efficiency obtained with HIJING 1.36 and DPMJet.
An additional 1.2\% uncertainty comes from the difference in the scaling factors due to the events without vertex using the two event generators, as discussed in Sec.~\ref{analysis}.
A Monte Carlo test was also carried out with DPMJet to check the difference in the results obtained from NSD generated events and from selected events, resulting in a difference of 0.2\% for the MB result, absorbed in the trigger efficiency uncertainty, and of 1.7\% (0.2\%) for 80--100\% (60--80\%) centrality bins. 
The contribution due to the subtraction of the background is studied using an alternative method where fake hits are injected into real events and it gives rise to a 0.3\% uncertainty.
The uncertainty from the material budget is 0.1\%, while the uncertainty due to the particle composition amounts to 0.3\%.
The contributions from the extrapolation down to zero $p_{\rm T}$ and from the pileup are found to be negligible.

The final systematic uncertainties assigned to the measurements are the quadratic sums of the individual contributions.
An overview of the systematic uncertainties is presented in Tab.~\ref{tab:systematics}.
For MB \dndetalab , they amount to 3.0\%. 
For centrality-dependent measurements the total uncertainty for central events is 2.6\%. 
For the most peripheral events it is 3.1\% at mid-rapidity and 3.6\% for $|\eta_{\rm lab}|=1.8$.
The difference in uncertainty between the MB and the centrality-dependent measurement is mostly due to the contributions from the selection efficiency for NSD, which are not included in the centrality-dependent measurement, and to the difference among \pPb\ and Pb--p collisions, which is more relevant for the most peripheral events at $|\eta_{\rm lab}|=1.8$.

\section{Results}
\label{res}

\begin{figure}[t]
\begin{center}
\includegraphics[width=\figlen]{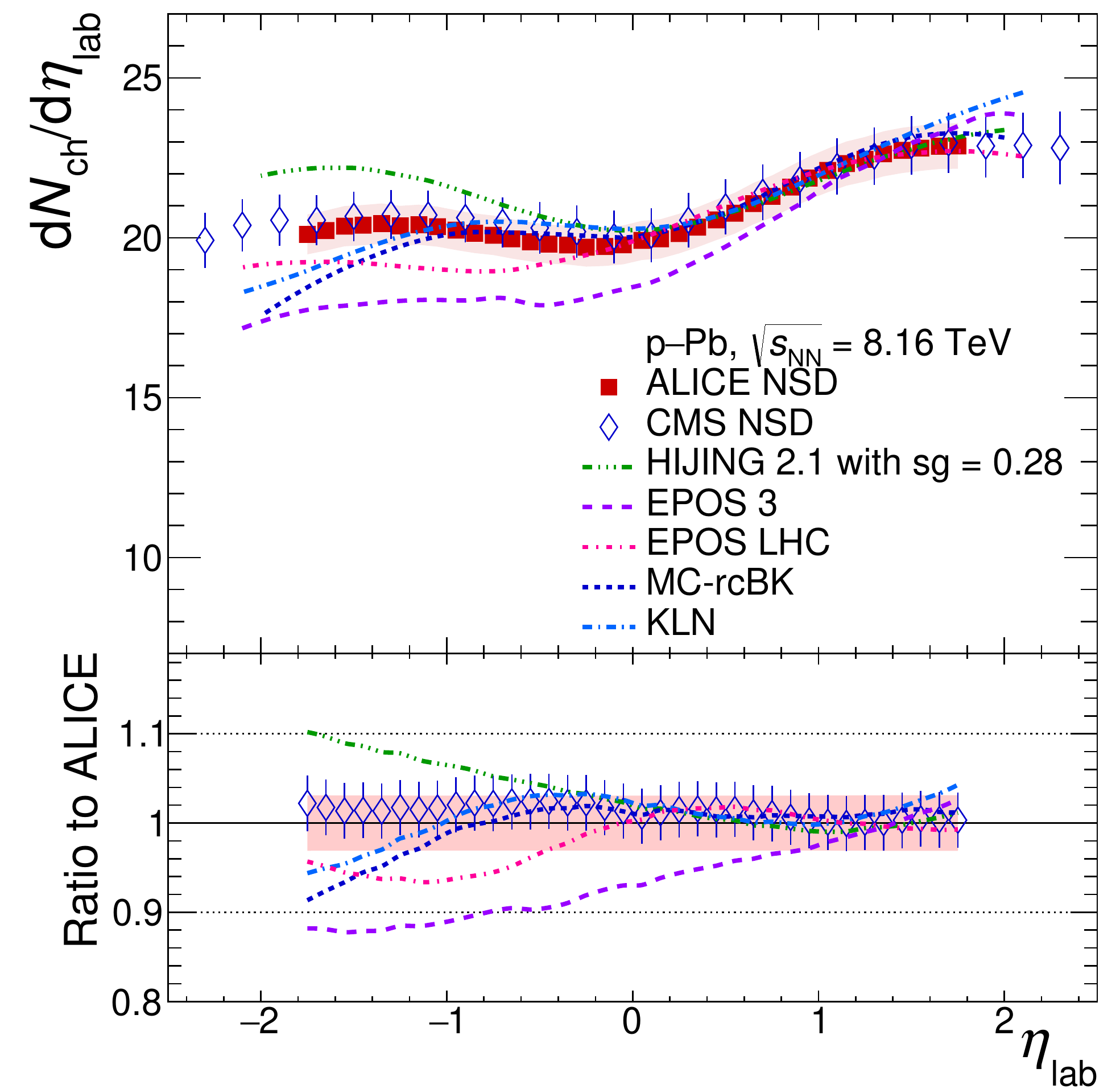}
\caption{Red squares show the measured pseudorapidity density of charged particles in \pPb\ NSD collisions at \snn\ $=8.16$ TeV in ALICE, with total systematic uncertainties shown as bands, compared with CMS results~\cite{Sirunyan:2017vpr} and theoretical predictions shifted to the laboratory system~\cite{Deng:2010mv, Pierog:2013ria, Drescher:2000ha, Albacete:2010ad, Kharzeev:2002ei}. The bottom panel shows the ratio to ALICE data.}
\label{fig:4}
\end{center}
\end{figure}
\begin{figure}[t]
\begin{center}
\includegraphics[width=\figlen]{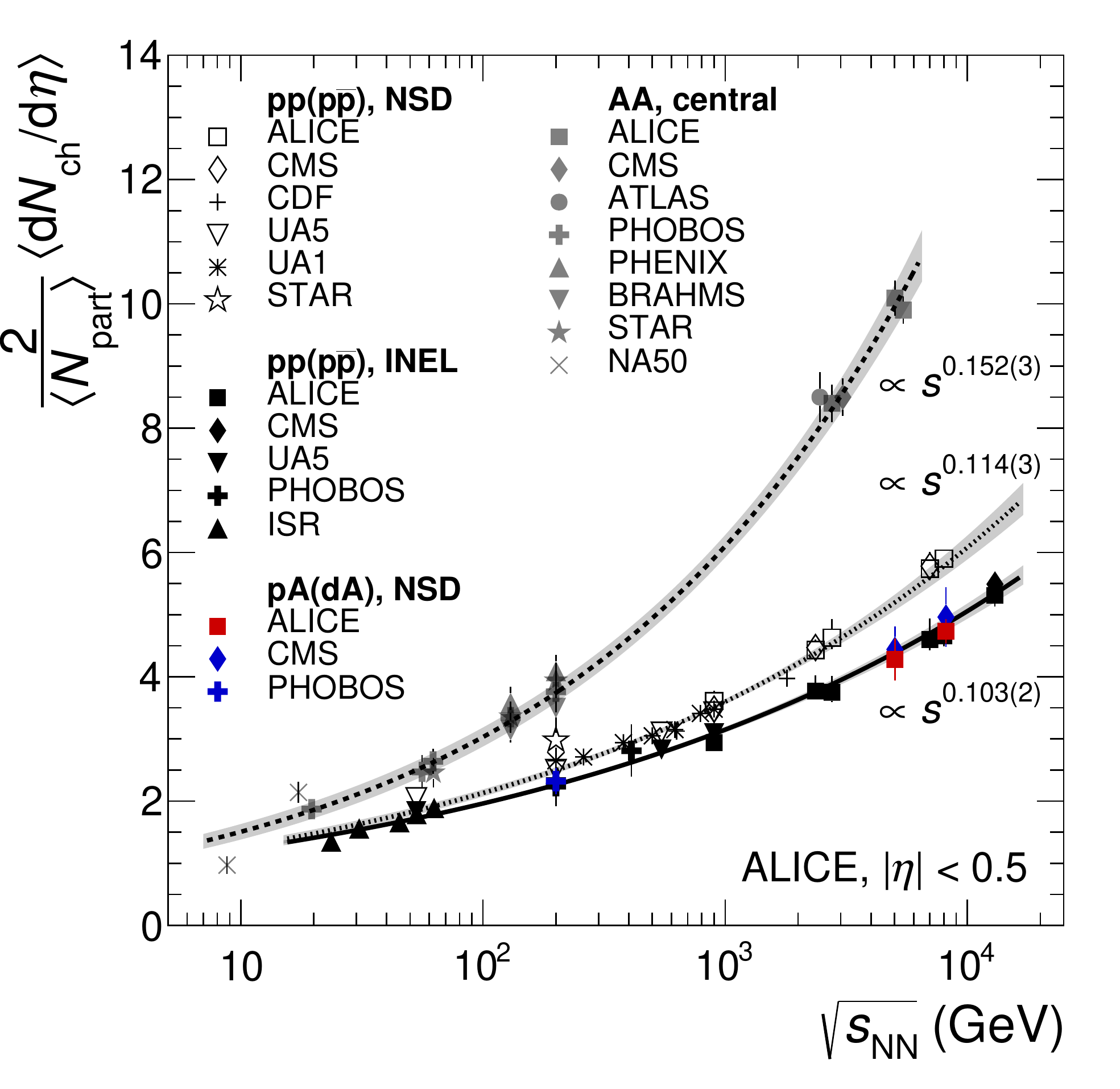}
\caption{Values of \dNdetape\ for pA~\cite{Sirunyan:2017vpr, ALICE:2012xs, Back:2003hx}, pp and \ppbar ~\cite{Thome:1977ky, Alpgard:1982kx, Alner:1986xu, Albajar:1989an, Abe:1989td, Khachatryan:2010xs, Khachatryan:2010us, Adam:2015gka,Khachatryan:2015jna,Adam:2015pza} along with those from central AA collisions~\cite{Acharya:2018hhy, Adam:2015ptt, Aamodt:2010cz, ATLAS:2011ag, Chatrchyan:2011pb, Abreu:2002fw, Bearden:2001xw, Bearden:2001qq, Adcox:2000sp, Alver:2010ck, Abelev:2008ab} as a function of \snn\ are shown, for $|\eta|<0.5$. 
All values of \avNpart\ used for normalisation of data are the results of Glauber model calculations.
The $s$-dependencies of the \pp\ (\ppbar) inelastic (INEL) and \pPb\ collisions data are proportional to $s_{\rm NN}^{0.103}$ (solid line), while pp (\ppbar ) NSD are proportional to $s_{\rm NN}^{0.114}$ (dashed middle line). AA are proportional to $s_{\rm NN}^{0.152}$ (dashed upper line). The bands show the uncertainties on the extracted power-law dependencies.}
\label{fig:3}
\end{center}
\end{figure}
\begin{figure}[t]
    \begin{subfigure}
    \centering
        \includegraphics[width=0.49\textwidth]{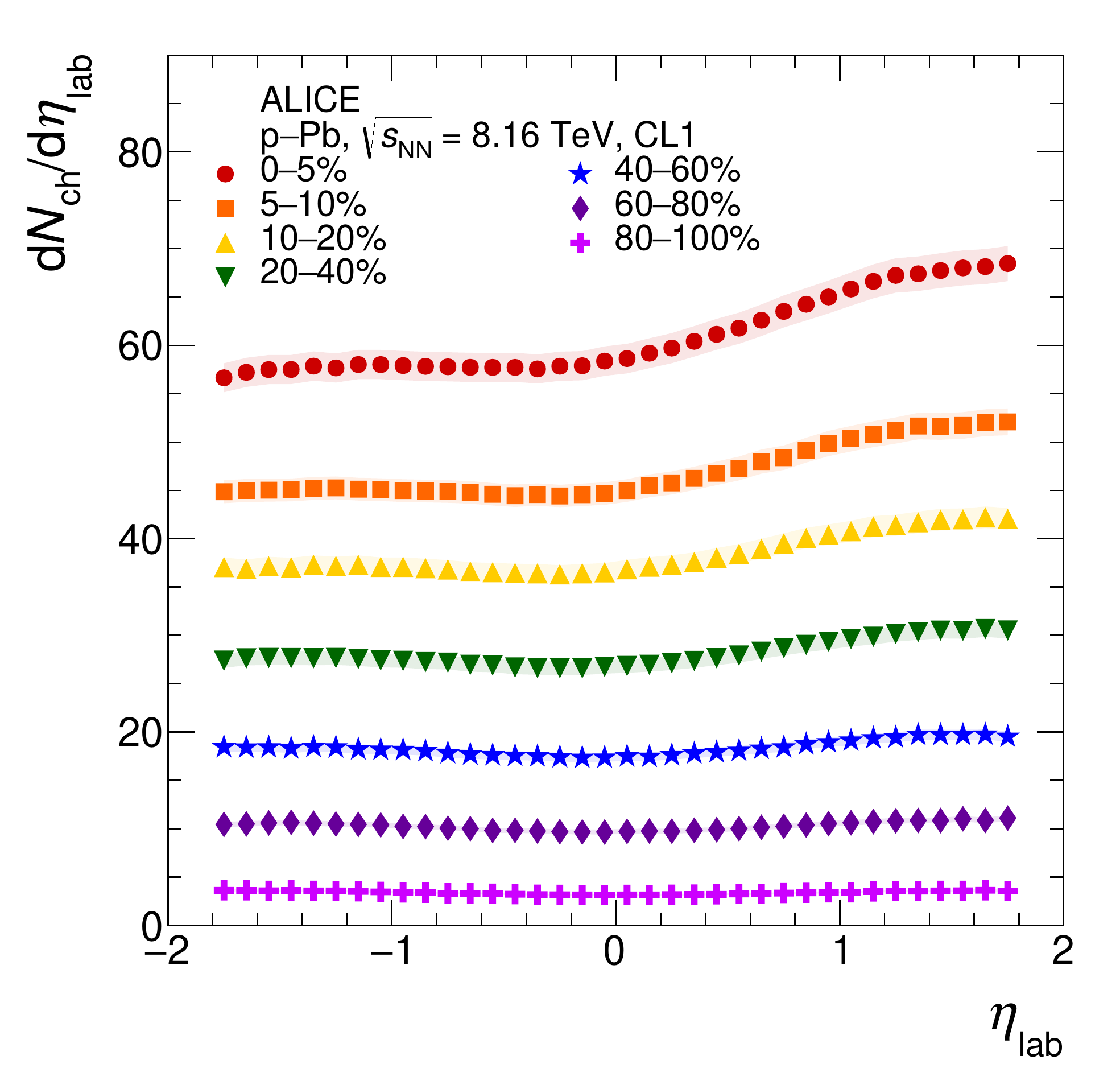}
    \end{subfigure}
	\begin{subfigure}
	\centering
        \includegraphics[width=0.49\textwidth]{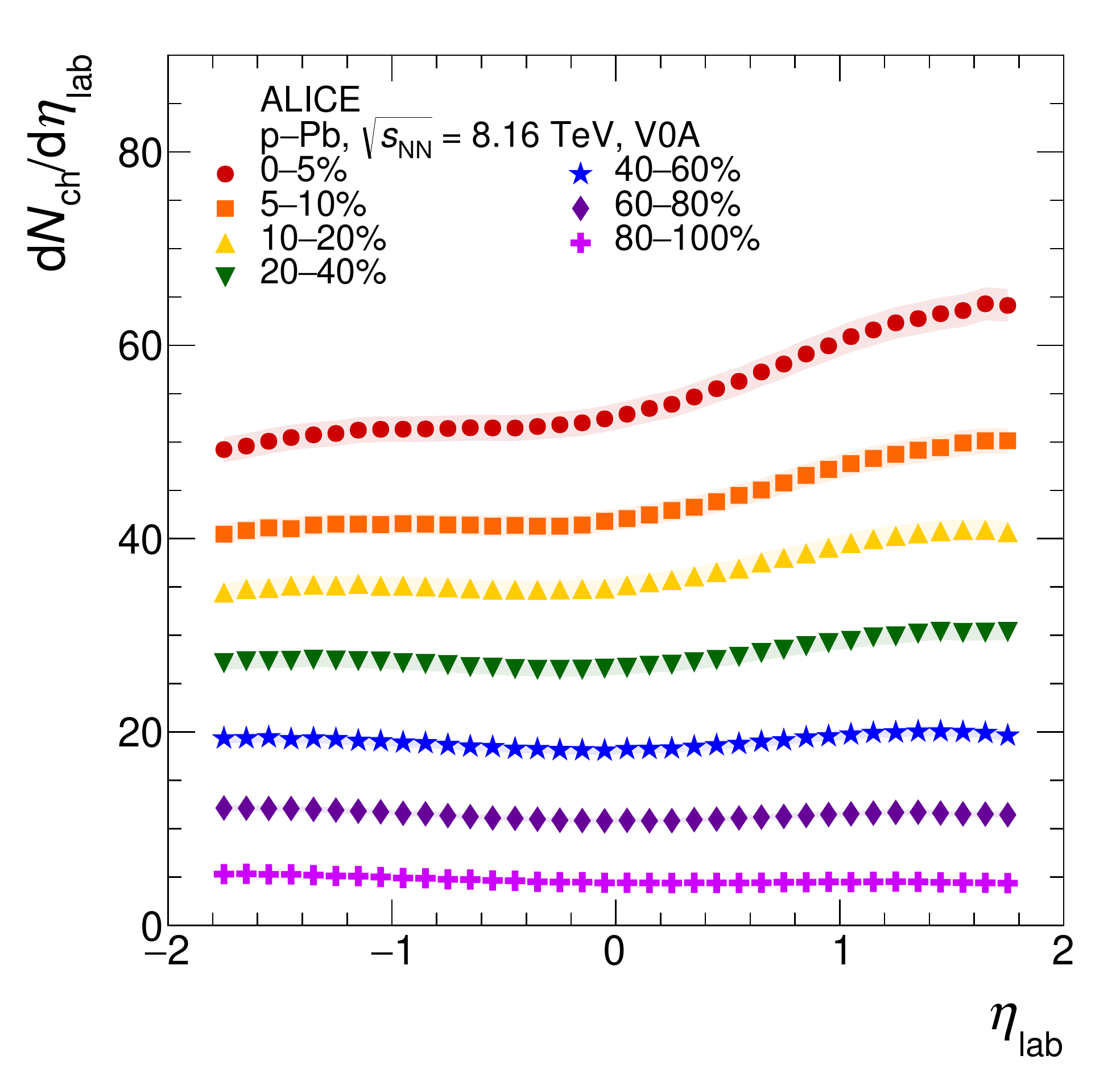}
	\end{subfigure} 
	\begin{subfigure}
	\centering
        \includegraphics[width=0.49\textwidth]{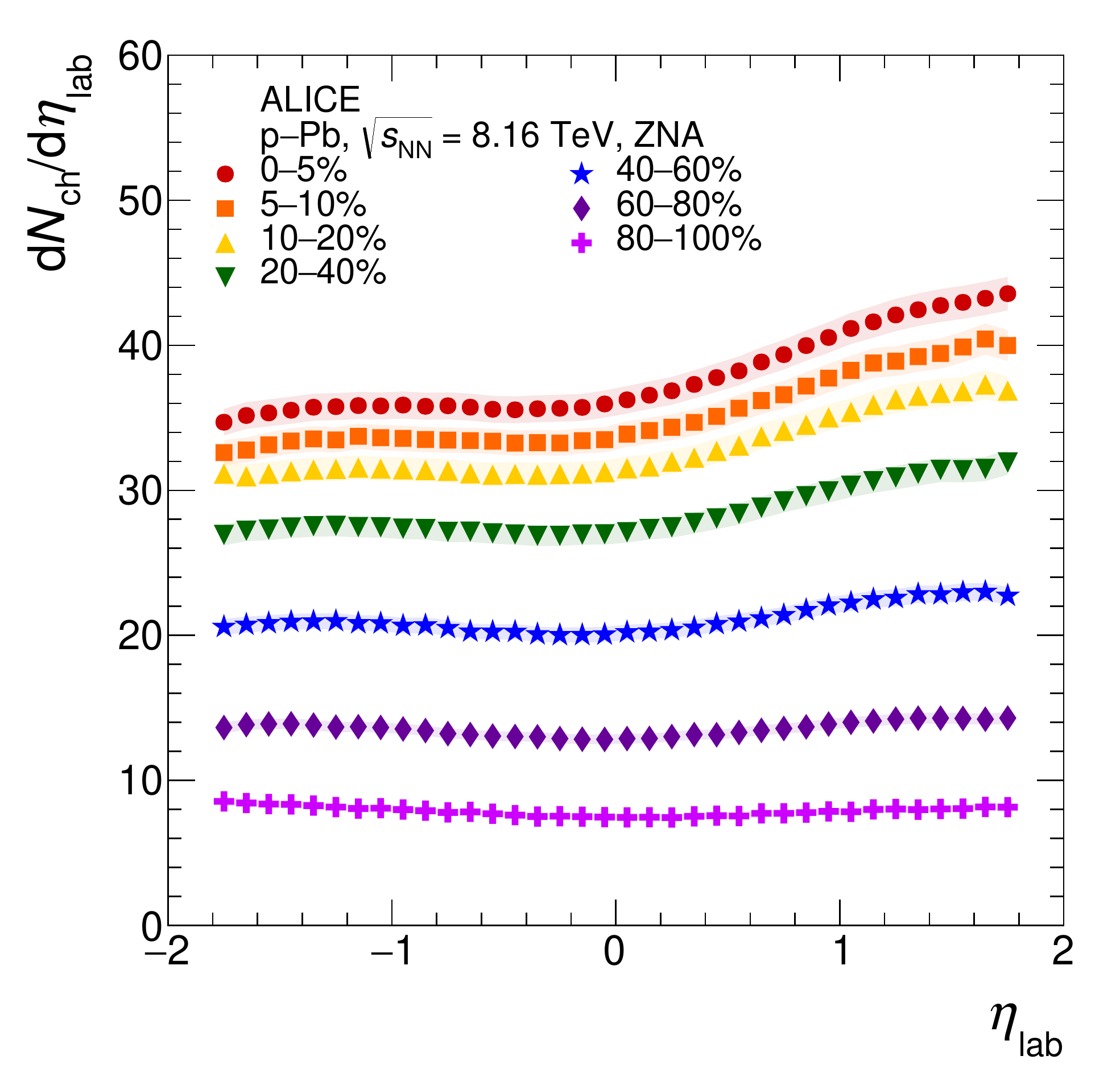}
	\end{subfigure} 
\caption{Pseudorapidity density of charged particles in \pPb\ NSD collisions at \snn\ $=8.16$ TeV for various centrality classes and estimators: CL1 (top left), V0A (top right) and ZNA (bottom left).}
\label{fig:5}
\end{figure}
\begin{figure}[t]
\begin{center}
    \begin{subfigure}
    \centering
        \includegraphics[width=0.49\textwidth]{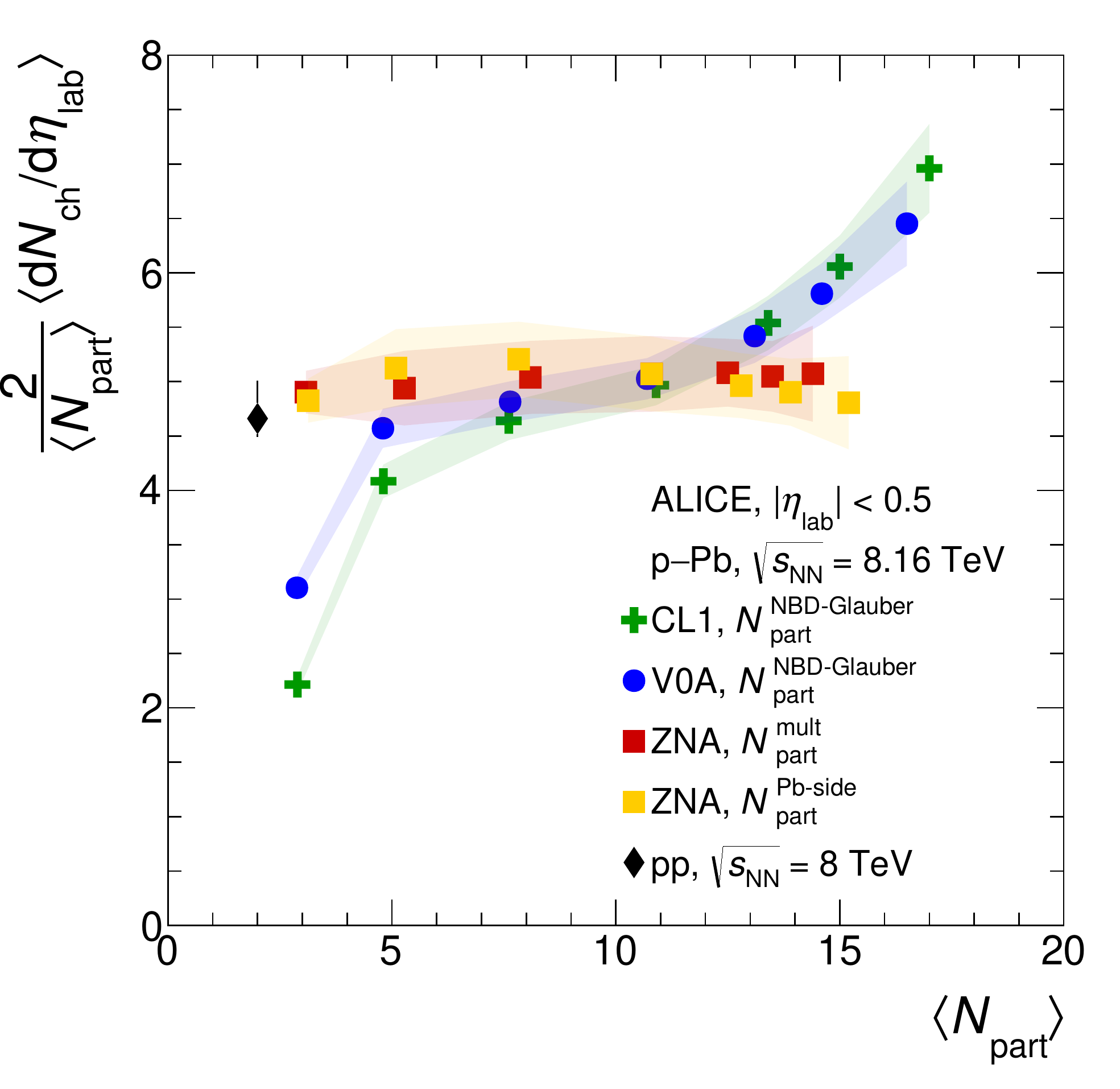}
    \end{subfigure}
    \begin{subfigure}
    \centering
        \includegraphics[width=0.49\textwidth]{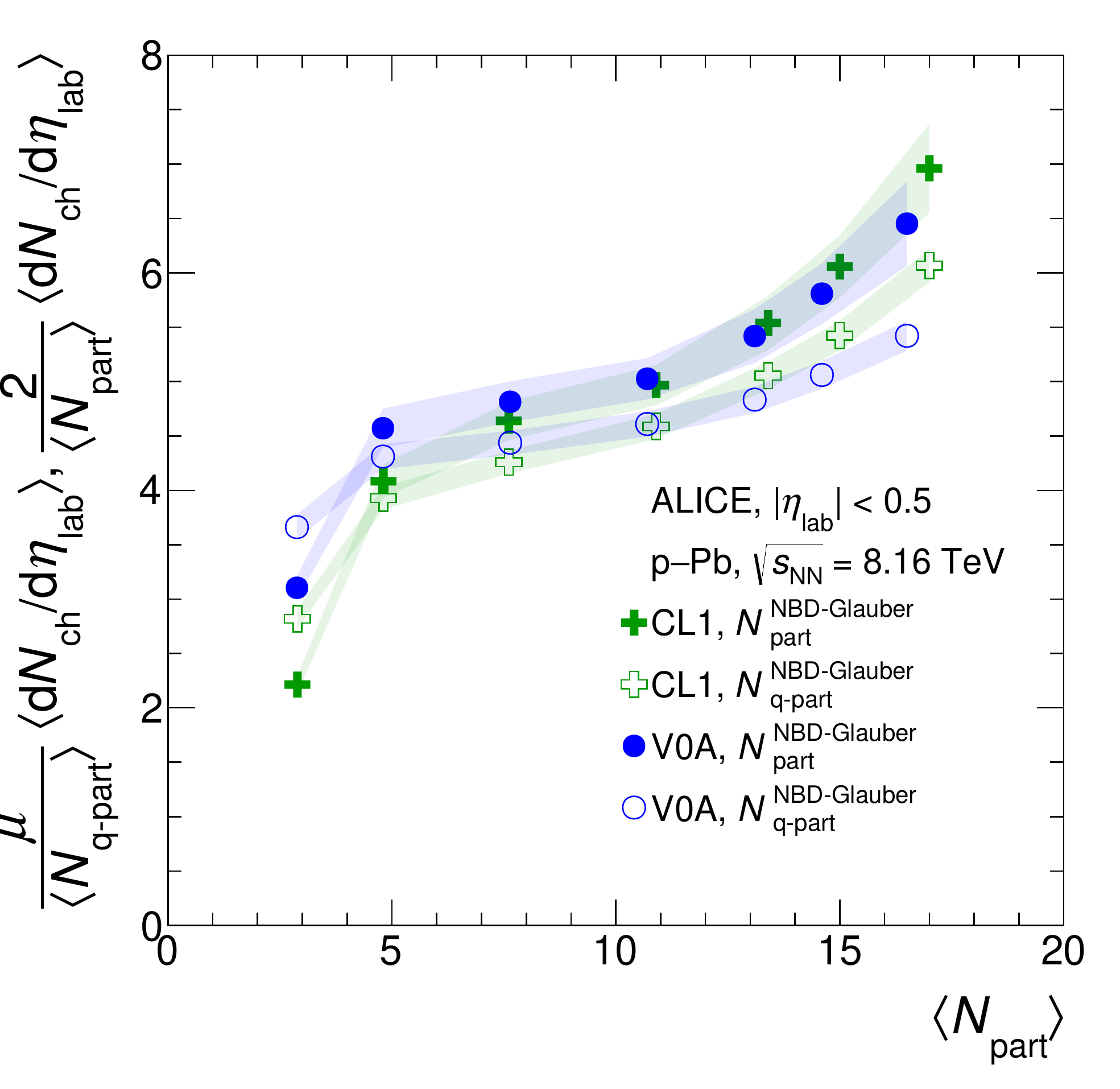}
	\end{subfigure} 
\caption{Left: \dNdetapelab\ in \pPb\ collisions at \snn\ $=8.16$ TeV and pp at 8 TeV~\cite{Adam:2015gka} as a function of \avNpart\ for different centrality estimators. Right: \dNdetaquarklab\ for $N_{\rm c}=5$, open points, with $\mu=4.44$.}
\label{fig:6}
\end{center}
\end{figure}

The pseudorapidity density as a function of $\eta_{\rm lab}$ is presented in Fig.~\ref{fig:4} for $|\eta_{\rm lab}|<1.8$.
An asymmetry between the proton and the lead hemispheres is observed, and the number of charged particles is higher in the Pb-going side (positive $\eta_{\rm lab}$).
The ALICE measurement is compared with the pseudorapidity density measured by CMS~\cite{Sirunyan:2017vpr} showing very good agreement within systematic uncertainties, although CMS results exclude prompt leptons. 
The result is also compared with several models with different descriptions of particle production, all shifted by $\eta_{\rm lab}=0.465$ to take into account the shift to the laboratory system.
In the improved HIJING 2.1~\cite{Deng:2010mv, Xu:2012au} version the Cronin effect is included, as well as a strong nuclear shadowing effect (sg $=0.28$) in order to explain the global properties of the final hadron system in \pPb\ collisions~\cite{ALICE:2012xs}.
The model describes well both the normalisation and the shape of the distribution for the Pb-going side, while it overestimates the p-going side, showing a symmetric behaviour, as for the \pPb\ collisions at 5.02 TeV.
The \dndetalab\ versus $\eta_{\rm lab}$ is compared with two different versions of EPOS.
EPOS LHC~\cite{Pierog:2013ria} is a tune of EPOS 1.99 based on LHC data.
It is designed to describe all bulk properties of hadronic interactions and based on Gribov-Regge theory for partons.
It incorporates collective effects with a separation of the initial state into a core and a corona.
EPOS LHC reproduces the Pb-going side, although it underestimates the p-going side of the distribution, showing a stronger asymmetry than data.
EPOS 1.99 contains collective flow parametrised at freeze-out, while EPOS 3~\cite{Drescher:2000ha, Werner:2010aa, Werner:2013tya} includes a full viscous hydrodynamical simulation.
It starts from flux tube initial conditions, which are generated in the Gribov-Regge multiple scattering framework.
It reproduces the most forward part of the distribution in the Pb-going side, but underestimates both the normalisation, the mid-rapidity part and the p-going side of the \dndetalab\ distribution.
Finally, the distribution is compared with two saturation-based models: MC-rcBK~\cite{Albacete:2010ad, Albacete:2012xq} and KLN~\cite{Kharzeev:2002ei, Dumitru:2011wq}, which contain a mechanism to limit the number of partons and particles produced.
The MC-rcBK results are obtained using the McLerran-Venugopalan model ($\gamma=1$)~\cite{McLerran:1993ni} for the Albacete-Armesto-Milhano-Quiroga-Salgado initial conditions~\cite{Albacete:2013ei}.
Saturation-based models are the ones which perform better, underlining the necessity of a mechanism to limit the number of partons produced. 
Indeed, both MC-rcBK and KLN reproduce the distribution well, within the uncertainties of data, and start to deviate in the region $\eta_{\rm lab}<-1.3$.
The MC-rcBK model better predicts the \pPb\ collisions at 8.16 TeV than the distribution at 5.02 TeV.
The shadowing mechanism used by HIJING is not sufficient to limit the partons produced in the p-going side.
Both EPOS and HIJING contain final-state effects, and the performance is worse than for models based on initial-state effects only, like MC-rcBK and KLN. 
This means that for the \dndeta\ observable final-state effects do not play a role, for the models considered.
Nevertheless, all models lie within about 10\% when compared with data, and reproduce within systematic uncertainties the Pb-going side.

The charged-particle pseudorapidity density in the laboratory system for $|\eta_{\rm lab}|<0.5$  is \dndetalab\ $=20.08$ $\pm$ 0.01 (stat.) $\pm$ 0.61 (syst.).
In the following, the statistical uncertainty is considered to be negligible.
The data are integrated in the range $-0.965<\eta_{\rm lab}<0.035$ and corrected for the effect of the rapidity shift to retrieve the \dndeta\ in the centre-of-mass system.
The correction for the pseudorapidity shift is estimated from HIJING 1.36~\cite{Wang:1991hta} to be $-3.7$\% $\pm$ 1.9\%.
The resulting pseudorapidity density in the centre of mass is \mbox{\dndeta\ $=19.1$ $\pm$ $0.7$}.

The charged-particle production is scaled by $N_{\rm part}/2$, calculated with a Glauber model as explained in Sec.~\ref{centrality}, in order to compare the bulk particle production in different collision systems.
The number of participants for MB events is $8.09$ $\pm$ $0.17$.
The value normalised to the number of participants divided by 2 gives \dndeta\ $\times(2/N_{\rm part})=4.73$ $\pm$ $0.20$.
In Fig.~\ref{fig:3}, this quantity is compared with lower energy \pPb\ measurements by ALICE~\cite{ALICE:2012xs} as well as by CMS~\cite{Sirunyan:2017vpr} and d--Au measurements at RHIC~\cite{Back:2003hx}, showing that the values overlap with \dndeta\ measurements for inelastic pp collisions~\cite{Adam:2015gka,Khachatryan:2015jna,Adam:2015pza}.
The dependence of \dNdeta\ on the centre-of-mass energy can be fitted with a power-law function of the form $\alpha\cdot s^{\beta}$. 
This gives an exponent, under the assumption of uncorrelated uncertainties, of $\beta = 0.103 \pm 0.002$. 
It is a much weaker $s$-dependence than for AA collisions~\cite{Acharya:2018hhy, Adam:2015ptt, Aamodt:2010cz, ATLAS:2011ag, Chatrchyan:2011pb, Abreu:2002fw, Bearden:2001xw, Bearden:2001qq, Adcox:2000sp, Alver:2010ck, Abelev:2008ab}, where a value of $\beta = 0.152 \pm 0.003$ is obtained. 
The fit results are plotted with their uncertainties shown as shaded bands. 
The result at \mbox{\snn\ = 8.16 TeV} confirms the trend established by lower energy data since the exponent $\beta$ is not significantly different when the new point is excluded from the fit.
The values for \pPb\ and \dAu\ collisions fall on the inelastic pp curve, indicating that the strong rise in AA might not be solely related to the multiple collisions undergone by the participants since the proton in pA collisions also encounters multiple nucleons.
As the contribution of diffractive processes to the selected \pPb\ sample is negligible, it is expected that the NSD and inelastic selection belong to the same curve for \pPb , and that this slope corresponds to the one obtained from the inelastic pp curve.

The pseudorapidity density as a function of $\eta_{\rm lab}$ is presented in Fig.~\ref{fig:5} for $|\eta_{\rm lab}|<1.8$ for different centrality intervals, from most central 0--5\% to most peripheral 80--100\% events.
The results for the CL1 estimator have a strong bias due to the complete overlap with the tracking region. 
V0A has a small multiplicity fluctuation bias due to the enhanced contribution from the Pb-fragmentation region. 
Finally, the ZNA measurement based on the energy deposited in the ZN does not have multiplicity bias.
The CL1 (ZNA) estimator produces the largest (lowest) values for the most central events and the lowest (largest) values for the most peripheral events.
It is worth noting that for all the estimators used to select centrality the asymmetry is evident for most central events, while the results for 60--80\% and 80--100\% classes,  where the \avNpart\ are around 4.5 and 3, respectively, are symmetric. 

The left panel of Fig.~\ref{fig:6} shows \dNdetapelab\ as a function of \avNpart\ for various centrality estimators. 
For CL1 and V0A the \avNpart\ from the Glauber model are used and the resulting \dNdetapelab\ has a steep increase for most central events (higher \avNpart ) due to the strong multiplicity bias discussed in Sec.~\ref{centrality}. 
The rise is steeper for CL1, where the overlap of the centrality selection region with the tracking region is maximal.
For the ZNA estimator, two sets of \avNpart\ are used corresponding to the two different hybrid method selections. 
For both $N_{\rm part}^{\rm mult}$ and $N_{\rm part}^{\text{Pb-side}}$ the trend is similar and extrapolates to the pp point at \s\ $=8$ TeV. 
The overall \avNpart\ dependence of \dNdetapelab\ for the ZNA estimator is flat and the \avNpart\ range is more limited when the selection is made in a well separated pseudorapidity region, rather than for multiplicity-based estimators (CL1 and V0A). 

A Glauber Monte Carlo calculation based on single quark scattering is also performed~\cite{Eremin:2003qn, Loizides:2016djv}, as it was done for AA collisions~\cite{Adam:2015ptt, Acharya:2018hhy}.
Quark constituents are located around the nucleon centre, where the proton density is modelled by a function of the proton radius. 
To account for effective partonic degrees of freedom, $N_{\rm c}=$ 5 quark constituents have been selected, since this number of constituents was tested for AA collisions and resulted in
a constant charged-particle production rate per constituent quark.
The effective inelastic cross section for constituent-quark collisions is set to 11.0 mb for 5 constituent quarks to match the 72.5 mb nucleon cross section for \pPb\ interactions at 8.16 TeV~\cite{Loizides:2017ack}.
The effective cross sections are constrained using nuclear reaction cross sections~\cite{Loizides:2016djv}.
The right panel of Fig.~\ref{fig:6} shows the \dNdetaquarklab\ scaled by the average number of participating quarks, $\mu$, in \pp\ collisions, which is 4.44 out of 10 participating quarks for $N_{\rm c}=5$, as a function of \Npart\ (open points).
For the multiplicity-based estimators, CL1 and V0A, there is an increase for the most central and decrease for the most peripheral events with a trend that resembles the one for \Npart\ scaling  (full points) but with decreased slope.
This fact suggests that nuclear-geometrical effects are represented in terms of constituent participant quarks, but not as well as observed for AA collisions~\cite{Adare:2015bua, Adam:2015ptt, Acharya:2018hhy}, meaning that the multiplicity-fluctuation bias might influence also the quark participants scaling.
The \dNdetaquarklab\ has been measured also for 3 constituent quarks, with an inelastic cross section of 22.5 mb and $\mu=3.54$, showing a distribution in between the \Npart\ and $N_{\text{q-part}}$ points.

\section{Summary and conclusions}
\label{summary}
Summarising, the charged-particle pseudorapidity density in $|\eta_{\rm lab}|<1.8$ in NSD \pPb\ collisions at \mbox{\snn\ $=8.16$ TeV} is presented.
A value of \dndeta\ $=19.1\pm0.7$ is measured at mid-rapidity, corresponding to $4.73\pm0.20$ charged particles per unit of pseudorapidity per participant pair, \avNpart$/2$, calculated with the Glauber model. 
The new measurement is 9.5\% higher than the value at \snn\ $=5.02$ TeV.
The dependence of \dNdeta\ on the centre-of-mass energy is fitted with a power-law function, which gives a much weaker $s$-dependence than for AA collisions.
The MB \dndetalab\ distribution as a function of $\eta_{\rm lab}$ is compared with CMS results, showing good agreement within uncertainties, and to different models: HIJING 2.1, EPOS (versions LHC and 3) and two saturation-based models, MC-rcBK and KLN.
All models can reproduce the data within about 10\%, which is a sound achievement given the complexity in describing soft-QCD processes.
The best performance comes from saturation-based models, and final-state effects seem not to improve the description of \dndeta .
Nevertheless, the results provide further constraints for models describing high-energy hadron collisions.
The pseudorapidity density for various centrality estimators has been shown and the asymmetry, typical of asymmetric collision systems like \pPb , is evident for most central events, while results for 60--80\% and 80--100\% centrality classes are symmetric. 
The methods to select centrality in \pPb\ collisions based on multiplicity measurements have been presented and they induce a multiplicity-fluctuation bias.
Results with a selection based on multiplicity estimators at mid-rapidity or within a few units of pseudorapidity and \avNpart\ from the Glauber model are lower for peripheral values of \dNdetape\ and higher for most central collisions than the pp value.
On the contrary, with centrality selected by the energy deposited in the ZDC, and assuming that the multiplicity in the Pb-going direction is proportional to $N_{\rm  part}^{\rm Pb-side}$, the overall behaviour of \dNdetape\ as a function of \avNpart\ is flat, and agrees with the pp measurement at 8 TeV.

%
%

\newenvironment{acknowledgement}{\relax}{\relax}
\begin{acknowledgement}
\section*{Acknowledgements}
The ALICE Collaboration would like to thank  J.~Albacete, W.-T.~Deng, A.~Dumitru, T.~Pierog and K.~Werner for helpful discussions on their model predictions.

The ALICE Collaboration would like to thank all its engineers and technicians for their invaluable contributions to the construction of the experiment and the CERN accelerator teams for the outstanding performance of the LHC complex.
The ALICE Collaboration gratefully acknowledges the resources and support provided by all Grid centres and the Worldwide LHC Computing Grid (WLCG) collaboration.
The ALICE Collaboration acknowledges the following funding agencies for their support in building and running the ALICE detector:
A. I. Alikhanyan National Science Laboratory (Yerevan Physics Institute) Foundation (ANSL), State Committee of Science and World Federation of Scientists (WFS), Armenia;
Austrian Academy of Sciences and Nationalstiftung f\"{u}r Forschung, Technologie und Entwicklung, Austria;
Ministry of Communications and High Technologies, National Nuclear Research Center, Azerbaijan;
Conselho Nacional de Desenvolvimento Cient\'{\i}fico e Tecnol\'{o}gico (CNPq), Universidade Federal do Rio Grande do Sul (UFRGS), Financiadora de Estudos e Projetos (Finep) and Funda\c{c}\~{a}o de Amparo \`{a} Pesquisa do Estado de S\~{a}o Paulo (FAPESP), Brazil;
Ministry of Science \& Technology of China (MSTC), National Natural Science Foundation of China (NSFC) and Ministry of Education of China (MOEC) , China;
Ministry of Science and Education, Croatia;
Centro de Aplicaciones Tecnol\'{o}gicas y Desarrollo Nuclear (CEADEN), Cubaenerg\'{\i}a, Cuba;
Ministry of Education, Youth and Sports of the Czech Republic, Czech Republic;
The Danish Council for Independent Research | Natural Sciences, the Carlsberg Foundation and Danish National Research Foundation (DNRF), Denmark;
Helsinki Institute of Physics (HIP), Finland;
Commissariat \`{a} l'Energie Atomique (CEA) and Institut National de Physique Nucl\'{e}aire et de Physique des Particules (IN2P3) and Centre National de la Recherche Scientifique (CNRS), France;
Bundesministerium f\"{u}r Bildung, Wissenschaft, Forschung und Technologie (BMBF) and GSI Helmholtzzentrum f\"{u}r Schwerionenforschung GmbH, Germany;
General Secretariat for Research and Technology, Ministry of Education, Research and Religions, Greece;
National Research, Development and Innovation Office, Hungary;
Department of Atomic Energy Government of India (DAE), Department of Science and Technology, Government of India (DST), University Grants Commission, Government of India (UGC) and Council of Scientific and Industrial Research (CSIR), India;
Indonesian Institute of Science, Indonesia;
Centro Fermi - Museo Storico della Fisica e Centro Studi e Ricerche Enrico Fermi and Istituto Nazionale di Fisica Nucleare (INFN), Italy;
Institute for Innovative Science and Technology , Nagasaki Institute of Applied Science (IIST), Japan Society for the Promotion of Science (JSPS) KAKENHI and Japanese Ministry of Education, Culture, Sports, Science and Technology (MEXT), Japan;
Consejo Nacional de Ciencia (CONACYT) y Tecnolog\'{i}a, through Fondo de Cooperaci\'{o}n Internacional en Ciencia y Tecnolog\'{i}a (FONCICYT) and Direcci\'{o}n General de Asuntos del Personal Academico (DGAPA), Mexico;
Nederlandse Organisatie voor Wetenschappelijk Onderzoek (NWO), Netherlands;
The Research Council of Norway, Norway;
Commission on Science and Technology for Sustainable Development in the South (COMSATS), Pakistan;
Pontificia Universidad Cat\'{o}lica del Per\'{u}, Peru;
Ministry of Science and Higher Education and National Science Centre, Poland;
Korea Institute of Science and Technology Information and National Research Foundation of Korea (NRF), Republic of Korea;
Ministry of Education and Scientific Research, Institute of Atomic Physics and Romanian National Agency for Science, Technology and Innovation, Romania;
Joint Institute for Nuclear Research (JINR), Ministry of Education and Science of the Russian Federation, National Research Centre Kurchatov Institute, Russian Science Foundation and Russian Foundation for Basic Research, Russia;
Ministry of Education, Science, Research and Sport of the Slovak Republic, Slovakia;
National Research Foundation of South Africa, South Africa;
Swedish Research Council (VR) and Knut \& Alice Wallenberg Foundation (KAW), Sweden;
European Organization for Nuclear Research, Switzerland;
National Science and Technology Development Agency (NSDTA), Suranaree University of Technology (SUT) and Office of the Higher Education Commission under NRU project of Thailand, Thailand;
Turkish Atomic Energy Agency (TAEK), Turkey;
National Academy of  Sciences of Ukraine, Ukraine;
Science and Technology Facilities Council (STFC), United Kingdom;
National Science Foundation of the United States of America (NSF) and United States Department of Energy, Office of Nuclear Physics (DOE NP), United States of America.    
\end{acknowledgement}

\bibliographystyle{utphys}   
\bibliography{dNdetapPb}

\newpage
\appendix
\section{The ALICE Collaboration}
\label{app:collab}

\begingroup
\small
\begin{flushleft}
S.~Acharya\Irefn{org140}\And 
F.T.-.~Acosta\Irefn{org20}\And 
D.~Adamov\'{a}\Irefn{org93}\And 
S.P.~Adhya\Irefn{org140}\And 
A.~Adler\Irefn{org74}\And 
J.~Adolfsson\Irefn{org80}\And 
M.M.~Aggarwal\Irefn{org98}\And 
G.~Aglieri Rinella\Irefn{org34}\And 
M.~Agnello\Irefn{org31}\And 
Z.~Ahammed\Irefn{org140}\And 
S.~Ahmad\Irefn{org17}
% ...
\And
\renewcommand\labelenumi{\textsuperscript{\theenumi}~}

\section*{Affiliation notes}
\renewcommand\theenumi{\roman{enumi}}
\begin{Authlist}
\item \Adef{org*}Deceased
\item \Adef{orgI}Dipartimento DET del Politecnico di Torino, Turin, Italy
\item \Adef{orgII}M.V. Lomonosov Moscow State University, D.V. Skobeltsyn Institute of Nuclear, Physics, Moscow, Russia
\item \Adef{orgIII}Department of Applied Physics, Aligarh Muslim University, Aligarh, India
\item \Adef{orgIV}Institute of Theoretical Physics, University of Wroclaw, Poland
\end{Authlist}

\section*{Collaboration Institutes}
\renewcommand\theenumi{\arabic{enumi}~}
\begin{Authlist}
\item \Idef{org1}A.I. Alikhanyan National Science Laboratory (Yerevan Physics Institute) Foundation, Yerevan, Armenia
\item \Idef{org2}Bogolyubov Institute for Theoretical Physics, National Academy of Sciences of Ukraine, Kiev, Ukraine
\item \Idef{org3}Bose Institute, Department of Physics  and Centre for Astroparticle Physics and Space Science (CAPSS), Kolkata, India
\item \Idef{org4}Budker Institute for Nuclear Physics, Novosibirsk, Russia
\item \Idef{org5}California Polytechnic State University, San Luis Obispo, California, United States
\item \Idef{org6}Central China Normal University, Wuhan, China
\item \Idef{org7}Centre de Calcul de l'IN2P3, Villeurbanne, Lyon, France
\item \Idef{org8}Centro de Aplicaciones Tecnol\'{o}gicas y Desarrollo Nuclear (CEADEN), Havana, Cuba
\item \Idef{org9}Centro de Investigaci\'{o}n y de Estudios Avanzados (CINVESTAV), Mexico City and M\'{e}rida, Mexico
\item \Idef{org10}Centro Fermi - Museo Storico della Fisica e Centro Studi e Ricerche ``Enrico Fermi', Rome, Italy
\item \Idef{org11}Chicago State University, Chicago, Illinois, United States
\item \Idef{org12}China Institute of Atomic Energy, Beijing, China
\item \Idef{org13}Chonbuk National University, Jeonju, Republic of Korea
\item \Idef{org14}Comenius University Bratislava, Faculty of Mathematics, Physics and Informatics, Bratislava, Slovakia
\item \Idef{org15}COMSATS Institute of Information Technology (CIIT), Islamabad, Pakistan
\item \Idef{org16}Creighton University, Omaha, Nebraska, United States
\item \Idef{org17}Department of Physics, Aligarh Muslim University, Aligarh, India
\item \Idef{org18}Department of Physics, Pusan National University, Pusan, Republic of Korea
\item \Idef{org19}Department of Physics, Sejong University, Seoul, Republic of Korea
\item \Idef{org20}Department of Physics, University of California, Berkeley, California, United States
\item \Idef{org21}Department of Physics, University of Oslo, Oslo, Norway
\item \Idef{org22}Department of Physics and Technology, University of Bergen, Bergen, Norway
\item \Idef{org23}Dipartimento di Fisica dell'Universit\`{a} 'La Sapienza' and Sezione INFN, Rome, Italy
\item \Idef{org24}Dipartimento di Fisica dell'Universit\`{a} and Sezione INFN, Cagliari, Italy
\item \Idef{org25}Dipartimento di Fisica dell'Universit\`{a} and Sezione INFN, Trieste, Italy
\item \Idef{org26}Dipartimento di Fisica dell'Universit\`{a} and Sezione INFN, Turin, Italy
\item \Idef{org27}Dipartimento di Fisica e Astronomia dell'Universit\`{a} and Sezione INFN, Bologna, Italy
\item \Idef{org28}Dipartimento di Fisica e Astronomia dell'Universit\`{a} and Sezione INFN, Catania, Italy
\item \Idef{org29}Dipartimento di Fisica e Astronomia dell'Universit\`{a} and Sezione INFN, Padova, Italy
\item \Idef{org30}Dipartimento di Fisica `E.R.~Caianiello' dell'Universit\`{a} and Gruppo Collegato INFN, Salerno, Italy
\item \Idef{org31}Dipartimento DISAT del Politecnico and Sezione INFN, Turin, Italy
\item \Idef{org32}Dipartimento di Scienze e Innovazione Tecnologica dell'Universit\`{a} del Piemonte Orientale and INFN Sezione di Torino, Alessandria, Italy
\item \Idef{org33}Dipartimento Interateneo di Fisica `M.~Merlin' and Sezione INFN, Bari, Italy
\item \Idef{org34}European Organization for Nuclear Research (CERN), Geneva, Switzerland
\item \Idef{org35}Faculty of Electrical Engineering, Mechanical Engineering and Naval Architecture, University of Split, Split, Croatia
\item \Idef{org36}Faculty of Engineering and Science, Western Norway University of Applied Sciences, Bergen, Norway
\item \Idef{org37}Faculty of Nuclear Sciences and Physical Engineering, Czech Technical University in Prague, Prague, Czech Republic
\item \Idef{org38}Faculty of Science, P.J.~\v{S}af\'{a}rik University, Ko\v{s}ice, Slovakia
\item \Idef{org39}Frankfurt Institute for Advanced Studies, Johann Wolfgang Goethe-Universit\"{a}t Frankfurt, Frankfurt, Germany
\item \Idef{org40}Gangneung-Wonju National University, Gangneung, Republic of Korea
\item \Idef{org41}Gauhati University, Department of Physics, Guwahati, India
\item \Idef{org42}Helmholtz-Institut f\"{u}r Strahlen- und Kernphysik, Rheinische Friedrich-Wilhelms-Universit\"{a}t Bonn, Bonn, Germany
\item \Idef{org43}Helsinki Institute of Physics (HIP), Helsinki, Finland
\item \Idef{org44}High Energy Physics Group,  Universidad Aut\'{o}noma de Puebla, Puebla, Mexico
\item \Idef{org45}Hiroshima University, Hiroshima, Japan
\item \Idef{org46}Hochschule Worms, Zentrum  f\"{u}r Technologietransfer und Telekommunikation (ZTT), Worms, Germany
\item \Idef{org47}Horia Hulubei National Institute of Physics and Nuclear Engineering, Bucharest, Romania
\item \Idef{org48}Indian Institute of Technology Bombay (IIT), Mumbai, India
\item \Idef{org49}Indian Institute of Technology Indore, Indore, India
\item \Idef{org50}Indonesian Institute of Sciences, Jakarta, Indonesia
\item \Idef{org51}INFN, Laboratori Nazionali di Frascati, Frascati, Italy
\item \Idef{org52}INFN, Sezione di Bari, Bari, Italy
\item \Idef{org53}INFN, Sezione di Bologna, Bologna, Italy
\item \Idef{org54}INFN, Sezione di Cagliari, Cagliari, Italy
\item \Idef{org55}INFN, Sezione di Catania, Catania, Italy
\item \Idef{org56}INFN, Sezione di Padova, Padova, Italy
\item \Idef{org57}INFN, Sezione di Roma, Rome, Italy
\item \Idef{org58}INFN, Sezione di Torino, Turin, Italy
\item \Idef{org59}INFN, Sezione di Trieste, Trieste, Italy
\item \Idef{org60}Inha University, Incheon, Republic of Korea
\item \Idef{org61}Institut de Physique Nucl\'{e}aire d'Orsay (IPNO), Institut National de Physique Nucl\'{e}aire et de Physique des Particules (IN2P3/CNRS), Universit\'{e} de Paris-Sud, Universit\'{e} Paris-Saclay, Orsay, France
\item \Idef{org62}Institute for Nuclear Research, Academy of Sciences, Moscow, Russia
\item \Idef{org63}Institute for Subatomic Physics, Utrecht University/Nikhef, Utrecht, Netherlands
\item \Idef{org64}Institute for Theoretical and Experimental Physics, Moscow, Russia
\item \Idef{org65}Institute of Experimental Physics, Slovak Academy of Sciences, Ko\v{s}ice, Slovakia
\item \Idef{org66}Institute of Physics, Homi Bhabha National Institute, Bhubaneswar, India
\item \Idef{org67}Institute of Physics of the Czech Academy of Sciences, Prague, Czech Republic
\item \Idef{org68}Institute of Space Science (ISS), Bucharest, Romania
\item \Idef{org69}Institut f\"{u}r Kernphysik, Johann Wolfgang Goethe-Universit\"{a}t Frankfurt, Frankfurt, Germany
\item \Idef{org70}Instituto de Ciencias Nucleares, Universidad Nacional Aut\'{o}noma de M\'{e}xico, Mexico City, Mexico
\item \Idef{org71}Instituto de F\'{i}sica, Universidade Federal do Rio Grande do Sul (UFRGS), Porto Alegre, Brazil
\item \Idef{org72}Instituto de F\'{\i}sica, Universidad Nacional Aut\'{o}noma de M\'{e}xico, Mexico City, Mexico
\item \Idef{org73}iThemba LABS, National Research Foundation, Somerset West, South Africa
\item \Idef{org74}Johann-Wolfgang-Goethe Universit\"{a}t Frankfurt Institut f\"{u}r Informatik, Fachbereich Informatik und Mathematik, Frankfurt, Germany
\item \Idef{org75}Joint Institute for Nuclear Research (JINR), Dubna, Russia
\item \Idef{org76}Korea Institute of Science and Technology Information, Daejeon, Republic of Korea
\item \Idef{org77}KTO Karatay University, Konya, Turkey
\item \Idef{org78}Laboratoire de Physique Subatomique et de Cosmologie, Universit\'{e} Grenoble-Alpes, CNRS-IN2P3, Grenoble, France
\item \Idef{org79}Lawrence Berkeley National Laboratory, Berkeley, California, United States
\item \Idef{org80}Lund University Department of Physics, Division of Particle Physics, Lund, Sweden
\item \Idef{org81}Nagasaki Institute of Applied Science, Nagasaki, Japan
\item \Idef{org82}Nara Women{'}s University (NWU), Nara, Japan
\item \Idef{org83}National and Kapodistrian University of Athens, School of Science, Department of Physics , Athens, Greece
\item \Idef{org84}National Centre for Nuclear Research, Warsaw, Poland
\item \Idef{org85}National Institute of Science Education and Research, Homi Bhabha National Institute, Jatni, India
\item \Idef{org86}National Nuclear Research Center, Baku, Azerbaijan
\item \Idef{org87}National Research Centre Kurchatov Institute, Moscow, Russia
\item \Idef{org88}Niels Bohr Institute, University of Copenhagen, Copenhagen, Denmark
\item \Idef{org89}Nikhef, National institute for subatomic physics, Amsterdam, Netherlands
\item \Idef{org90}NRC Kurchatov Institute IHEP, Protvino, Russia
\item \Idef{org91}NRNU Moscow Engineering Physics Institute, Moscow, Russia
\item \Idef{org92}Nuclear Physics Group, STFC Daresbury Laboratory, Daresbury, United Kingdom
\item \Idef{org93}Nuclear Physics Institute of the Czech Academy of Sciences, \v{R}e\v{z} u Prahy, Czech Republic
\item \Idef{org94}Oak Ridge National Laboratory, Oak Ridge, Tennessee, United States
\item \Idef{org95}Ohio State University, Columbus, Ohio, United States
\item \Idef{org96}Petersburg Nuclear Physics Institute, Gatchina, Russia
\item \Idef{org97}Physics department, Faculty of science, University of Zagreb, Zagreb, Croatia
\item \Idef{org98}Physics Department, Panjab University, Chandigarh, India
\item \Idef{org99}Physics Department, University of Jammu, Jammu, India
\item \Idef{org100}Physics Department, University of Rajasthan, Jaipur, India
\item \Idef{org101}Physikalisches Institut, Eberhard-Karls-Universit\"{a}t T\"{u}bingen, T\"{u}bingen, Germany
\item \Idef{org102}Physikalisches Institut, Ruprecht-Karls-Universit\"{a}t Heidelberg, Heidelberg, Germany
\item \Idef{org103}Physik Department, Technische Universit\"{a}t M\"{u}nchen, Munich, Germany
\item \Idef{org104}Research Division and ExtreMe Matter Institute EMMI, GSI Helmholtzzentrum f\"ur Schwerionenforschung GmbH, Darmstadt, Germany
\item \Idef{org105}Rudjer Bo\v{s}kovi\'{c} Institute, Zagreb, Croatia
\item \Idef{org106}Russian Federal Nuclear Center (VNIIEF), Sarov, Russia
\item \Idef{org107}Saha Institute of Nuclear Physics, Homi Bhabha National Institute, Kolkata, India
\item \Idef{org108}School of Physics and Astronomy, University of Birmingham, Birmingham, United Kingdom
\item \Idef{org109}Secci\'{o}n F\'{\i}sica, Departamento de Ciencias, Pontificia Universidad Cat\'{o}lica del Per\'{u}, Lima, Peru
\item \Idef{org110}Shanghai Institute of Applied Physics, Shanghai, China
\item \Idef{org111}St. Petersburg State University, St. Petersburg, Russia
\item \Idef{org112}Stefan Meyer Institut f\"{u}r Subatomare Physik (SMI), Vienna, Austria
\item \Idef{org113}SUBATECH, IMT Atlantique, Universit\'{e} de Nantes, CNRS-IN2P3, Nantes, France
\item \Idef{org114}Suranaree University of Technology, Nakhon Ratchasima, Thailand
\item \Idef{org115}Technical University of Ko\v{s}ice, Ko\v{s}ice, Slovakia
\item \Idef{org116}Technische Universit\"{a}t M\"{u}nchen, Excellence Cluster 'Universe', Munich, Germany
\item \Idef{org117}The Henryk Niewodniczanski Institute of Nuclear Physics, Polish Academy of Sciences, Cracow, Poland
\item \Idef{org118}The University of Texas at Austin, Austin, Texas, United States
\item \Idef{org119}Universidad Aut\'{o}noma de Sinaloa, Culiac\'{a}n, Mexico
\item \Idef{org120}Universidade de S\~{a}o Paulo (USP), S\~{a}o Paulo, Brazil
\item \Idef{org121}Universidade Estadual de Campinas (UNICAMP), Campinas, Brazil
\item \Idef{org122}Universidade Federal do ABC, Santo Andre, Brazil
\item \Idef{org123}University College of Southeast Norway, Tonsberg, Norway
\item \Idef{org124}University of Cape Town, Cape Town, South Africa
\item \Idef{org125}University of Houston, Houston, Texas, United States
\item \Idef{org126}University of Jyv\"{a}skyl\"{a}, Jyv\"{a}skyl\"{a}, Finland
\item \Idef{org127}University of Liverpool, Liverpool, United Kingdom
\item \Idef{org128}University of Science and Techonology of China, Hefei, China
\item \Idef{org129}University of Tennessee, Knoxville, Tennessee, United States
\item \Idef{org130}University of the Witwatersrand, Johannesburg, South Africa
\item \Idef{org131}University of Tokyo, Tokyo, Japan
\item \Idef{org132}University of Tsukuba, Tsukuba, Japan
\item \Idef{org133}Universit\'{e} Clermont Auvergne, CNRS/IN2P3, LPC, Clermont-Ferrand, France
\item \Idef{org134}Universit\'{e} de Lyon, Universit\'{e} Lyon 1, CNRS/IN2P3, IPN-Lyon, Villeurbanne, Lyon, France
\item \Idef{org135}Universit\'{e} de Strasbourg, CNRS, IPHC UMR 7178, F-67000 Strasbourg, France, Strasbourg, France
\item \Idef{org136} Universit\'{e} Paris-Saclay Centre d¿\'Etudes de Saclay (CEA), IRFU, Department de Physique Nucl\'{e}aire (DPhN), Saclay, France
\item \Idef{org137}Universit\`{a} degli Studi di Foggia, Foggia, Italy
\item \Idef{org138}Universit\`{a} degli Studi di Pavia, Pavia, Italy
\item \Idef{org139}Universit\`{a} di Brescia, Brescia, Italy
\item \Idef{org140}Variable Energy Cyclotron Centre, Homi Bhabha National Institute, Kolkata, India
\item \Idef{org141}Warsaw University of Technology, Warsaw, Poland
\item \Idef{org142}Wayne State University, Detroit, Michigan, United States
\item \Idef{org143}Westf\"{a}lische Wilhelms-Universit\"{a}t M\"{u}nster, Institut f\"{u}r Kernphysik, M\"{u}nster, Germany
\item \Idef{org144}Wigner Research Centre for Physics, Hungarian Academy of Sciences, Budapest, Hungary
\item \Idef{org145}Yale University, New Haven, Connecticut, United States
\item \Idef{org146}Yonsei University, Seoul, Republic of Korea
\end{Authlist}
\endgroup
\end{document}